\documentclass[12pt]{article}
\usepackage{a4wide}
\usepackage{amssymb}
\usepackage{graphicx}
\usepackage{xcolor}
\begin{document}
{\renewcommand{\thefootnote}{\fnsymbol{footnote}}
\begin{center}
{\LARGE Space-time superpositions as fluctuating geometries}\\
\vspace{1.5em}
K\"allan Berglund,$^1$\footnote{e-mail address: {\tt kallan\_berglund@alumni.brown.edu}}
Martin Bojowald,$^1$\footnote{e-mail address: {\tt bojowald@psu.edu}}
Aurora Colter$^1$\footnote{e-mail address: {\tt aurora.colter@gmail.com}}
and Manuel D\'{\i}az$^2$\footnote{e-mail address: {\tt manueldiaz@umass.edu}}
\\
\vspace{0.5em}
$^1$Institute for Gravitation and the Cosmos,\\
The Pennsylvania State
University,\\
104 Davey Lab, University Park, PA 16802, USA\\
\vspace{0.5em}
$^2$ Amherst Center for Fundamental Interactions,\\ Department of Physics, University of Massachusetts Amherst,\\ 426 Lederle Graduate Research Tower, Amherst, MA 01003 USA
\vspace{1.5em}
\end{center}
}

\setcounter{footnote}{0}

\begin{abstract}
Superpositions of black holes can be described geometrically using a combined canonical formulation for space-time and quantum states. A previously introduced black-hole model that includes quantum fluctuations of metric components is shown here to give full access to the corresponding space-time geometry of weak-field gravity in terms of suitable line elements with quantum corrections. These results can be interpreted as providing covariant formulations of the gravitational force implied by a distribution of black holes in superposition. They can also be understood as a distribution of quantum matter constituents in superposition for a single black hole. A detailed analysis in the weak-field limit reveals quantum corrections to Newton's potential in generic semiclassical states, as well as new bounds on quantum fluctuations, implied by the covariance condition, rather than the usual uncertainty principle. These results provide additional control on quantum effects in Newton's potential that can be used in a broad range of predictions to be compared with observations.
\end{abstract}

\section{Introduction}

A quantum state formed by a superposition of different massive objects is expected to generate a gravitational force dependent on quantum effects. Detailed derivations and studies of the resulting implications on test objects, which may themselves be in non-trivial quantum states, require ingredients from gravitational as well as quantum physics. Given the current lack of a complete and consistent quantum theory of gravity, only partial answers can be given to the question of how quantum test objects behave in a superposition state of different masses. Nevertheless, this setting is promising because it does not necessarily require high curvature, where a detailed theory of quantum gravity is believed to be essential, and can likely be addressed by semiclassical or other approximations.

\subsection{Current approaches}

Recent developments in quantum information methods applied to gravitating states and quantum reference frames have led to strong interest in the possible
implications of superpositions of states that experience or generate
gravitational fields. For instance, various proposals in
\cite{ProperTimeInterfer,ProperTimeInterfer2,ClassQuantProperTime,QuantumSwitchEarth}
analyze potential aspects of quantum superpositions of test-particle states at
locations that experience different values of proper time, due to their location in the gravitational potential well of a larger mass. In these and other cases, a classical background  gravitational field or a classical space-time geometry are assumed, usually at weak fields, which is experienced in different ways by
observers or quantum test particles depending on their states. For example, particles moving along the two arms of an interferometer experience different time dilations, if they travel at slightly displaced altitudes. Since the
combined superposition state simultaneously experiences two different values
of time dilation, there may be characteristic signatures in the interference
pattern that differ from the pattern in the absence of gravity. An analysis of the resulting state suggests the possibility to study gravitational effects in a quantum context, potentially using new experiments.

Models based on a classical background space-time are feasible because they do
not directly involve quantum gravity or quantum space-time, even if quantum
states are used for test objects. A related question closer to quantum aspects
of the gravitational field itself is how test objects might experience the
gravitational force generated by two (or more) masses in a quantum
superposition. Recent examples of such studies include
\cite{BHMassSup,QuantumSwitchSphSymm} for superpositions of spherically symmetric masses or shells.  In a first approximation, the question may again be analyzed in a weak-field context, using superpositions of Newtonian potentials. However, even though a Newtonian description is not necessarily relativistic, geometrical properties of space-time are used in the definition of proper time that determines how the internal degrees of freedom of a quantum test particle evolve at its position. This is particularly relevant to one of our results, as it involves deriving quantum corrections to the Newtonian potential.

The concept of proper time requires an understanding of space-time geometry because it is derived from a line element. In \cite{QuantumSupSpaceTime,QuantumSupFalling,QuantumSupConformal}, proper time experienced by an object in the gravitational field generated by a superposition of two masses was defined using the individual space-time geometries of each mass, and applying transformations between quantum reference frames to combine the individual classical-type results into a time duration experienced in a superposition of the two geometries. However, there was no space-time geometry that could describe the combined implication of both masses. From a fundamental perspective, such a description is incomplete because it is unable to describe important gravitational phenomena such as gravitational waves: If the two masses in superposition are orbiting around each other, they emit gravitational waves. Since these waves are produced by the combined system of both masses, they cannot be formulated as excitations on each of these individual space-times for a single mass; the system must be considered in a combined space-time description. Such studies may be sufficient to derive a specific kind of new, and potentially observable, effect related to the behavior of test objects. However, they do not show whether (or how) space-time geometry around a superposition of masses can be used for a complete description of gravitational phenomena. Field-theoretical aspects of gravity and general covariance require a consistent geometrical formulation of space-time, on which excitations such as gravitational waves can propagate. An open question is whether such a formulation can retain its classical form of Riemannian geometry if it is generated by superposition states. It is possible that geometrical concepts eventually have to be generalized in a suitable way to create such a formalism.

\subsection{Geometrical space-time structures}

A key issue in these endeavors is therefore whether a consistent geometrical structure can be derived from a superposition of two gravitational fields, and how to do so, if it is possible. The existence of basic geometrical concepts such as line elements or suitable generalizations compatible with a superposition state is also required. This is necessary for a complete description that makes it possible to extend the analysis to strong fields. In such cases, gravity is understood through general relativity as an implication of space-time curvature. The concept of space-time curvature, in turn, requires a meaningful definition of space-time geometry, tensor calculus, and general covariance. However, it is currently unclear what the superposition of two geometries should be like in a geometrical setting. For instance, is the superposition of two Riemannian geometries (described by space-time line elements) still Riemannian, or do we need a more general concept of geometry? If the superposition is still Riemannian, how do we derive a valid line element which faithfully describes geometrical properties of the superposition? Compared to the motion of quantum test masses on a given background space-time, these questions are much closer to fundamental issues in quantum gravity. Answering them could potentially suggest new classes of physical phenomena, such as, unique properties of gravitational waves produced by a black hole (in a superposition of the massive constituents of a progenitor star), or new behaviors of Hawking radiation.

Even in weak fields, a geometrical picture of superpositions of gravitational states is expected to have important distinctions. Superpositions of masses with a Newtonian gravitational potential are restricted by standard conditions of quantum mechanics, such as the uncertainty relation. Well-defined space-time geometries, and their potential superpositions, are also subject to the important condition of general covariance. This symmetry, expressed as 4-dimensional coordinate invariance or the freedom to choose arbitrary spacelike slicings of the space-time manifold, imposes restrictions on possible theories. One standard application is the classification of possible modifications of general relativity. For instance, the properties of curvature tensors and their invariants determine the form of possible higher-curvature effective actions. Similarly, general covariance is expected to restrict possible superpositions of gravitational fields, even if they are applied only in a weak-field limit. It would be impossible to notice such restrictions from general covariance, if one does not attempt to construct a geometrical space-time picture of superposed Newtonian potentials. In this context, the question posed here is whether there is an action principle or some other fundamental description that has superpositions of gravitational fields among its solutions. This important field-theoretical question cannot be addressed by current formulations of quantum reference frames, which focus on the properties of test objects.

General covariance is usually difficult to control if one starts with a non-covariant formulation, such as a spatially dependent potential. As is well known from decades of research on quantum gravity, the problem becomes even more challenging if quantum effects are included. For instance, a path-integral formulation may formally work with space-time tensors, but could
still violate general covariance if there are anomalies in the measure used to integrate over the space of metrics. Superpositions of masses lead to
non-Gaussian states on a field space, which may imply additional challenges
for common evaluations of path integrals based on saddle-point approximations. It is therefore important to develop a direct and tractable approach to analyze covariance conditions and their potential implications. A possible covariant approach has been developed in \cite{ScalarFieldCurv,DistCorr} and used for an analysis of space-time superpositions in \cite{BHMassSup}. However, in this case, the space-time geometry follows indirectly from the evolution of matter fields, based on field correlators. Here, we require that a fundamental description of gravitational superposition states be independent of matter properties and also work in vacuum situations. It should therefore be directly applied to the geometry of space-time, described by the metric or an alternative mathematical object.

It might seem that a full-fledged theory of quantum gravity is required to address these questions about the quantum properties of the gravitational field. However, we will show here that this is not the case. The situation is similar to derivations of quantum corrections to Newton's potential \cite{EffectiveNewton}, based on perturbative quantum gravity as an effective field theory \cite{EffectiveGR,BurgessLivRev}. These results are obtained from Feynman expansions. Alternatively, they can be calculated using path integrals. This effectively traces out quantum degrees of freedom such as fluctuations, quantum correlations, or higher moments of an underlying state. Quantum corrections obtained in this way depend on dynamical relationships between the moments of a state and expectation values of basic field operators, but they do not use explicit expressions for these moments. In a quantum mechanical context, these methods, described for this setting in \cite{EffAcQM}, can be extended to a more general treatment \cite{EffAc,Karpacz} in which moments and expectation values are initially independent and subject to equations of motion. An adiabatic approximation can then be used to solve for the moments in terms of expectation values. Inserting these solutions back into the equations of motion for expectation values is equivalent to the path-integral effective action from \cite{EffAcQM}. The generalization lies in the fact that it is not necessary to perform the adiabatic approximation. Non-adiabatic properties may indeed be important in a gravitational context, even for static backgrounds. This is due to the fact that a space-time covariant treatment implies non-adiabatic behavior if there are relevant variations of the fields, not only in time, but also in space.  Moreover, the general methods of \cite{EffAc,Karpacz} provide direct information about properties of  moments which may lead to additional physics insights. For instance, we will derive new bounds on quantum fluctuations of static black holes in addition to standard uncertainty relations.

\subsection{Combined geometry of space-time and quantum physics}

Importantly, these generalizations also allow us to consider non-Gaussian and possibly mixed states. In the present context, we are looking for superposition states of masses. In a Newtonian picture, such states could be understood as  wave functions in an energy representation of a mass superposition, or in a position representation describing the location of the superposed masses. However, masses or positions are not necessarily the fundamental degrees of freedom of a quantum theory of gravity in which a geometrical superposition state may be constructed. The precise nature of fundamental degrees of freedom depends on the approach to quantum gravity, but it is to be expected that a rather simple state for the superposition of two masses or two positions can only be obtained after a complicated procedure of tracing out infinitely many local degrees of freedom of a fundamental theory of quantum gravity. It is therefore important that physical results do not depend on assumptions of Gaussianity, purity, or other properties of states that are being superposed. The only reliable assumption is semiclassicality in a general sense, such as the existence of an expansion in $\hbar$, in order to preserve the classical limit in a Newtonian regime. A general parameterization of states by moments up to a certain order gives us enough freedom to obey this condition. As we will see, specific solutions for moments can be derived from geometrical constraints and staticity conditions.

We use these general methods in order to propose a new approach to the
question of superposition geometries based on a canonical formulation of
general relativity. This procedure has two main advantages in the present context. First, while the usual canonical formulation of general relativity does not work with space-time tensors and is not manifestly covariant, it implies strict algebraic conditions on possible equations of motion, which are compatible with general covariance. These conditions can be described by the requirement that the Poisson brackets between the local generators of space and time translations remain closed. These generators are formally given by the diffeomorphism and Hamiltonian constraints of canonical gravity, respectively. In theories known to be generally covariant, such as general relativity, evaluating canonical equations is usually considered more tedious than evaluating equations for space-time tensors, such as Einstein's equation. However, canonical equations have the great advantage of allowing us to analyze the covariance of proposed modifications of general relativity without having to know beforehand that the modifications are covariant, or whether they may be related to non-Riemannian geometries. Canonical gravity is therefore a useful tool for analyzing possible geometrical descriptions of superpositions of gravitational fields.

Secondly, canonical methods are useful because they employ phase-space
formulations of the gravitational field, which can be combined with suitable
phase-space formulations of quantum mechanics, such as the canonical effective methods introduced in \cite{EffAc,Karpacz}. General covariance is implemented in this picture by having a suitable form of gauge generators, as phase-space functions that obey certain algebraic relations based on the Poisson
bracket. One of the gauge generators is closely related to the gravitational
Hamiltonian because the Hamiltonian generates time translations, which are one example of covariant transformationa. The quantum Hamiltonian, written in a phase-space formulation of quantum mechanics, then provides a candidate for possible quantum modifications of the gauge generators. For instance, one can add fluctuation terms derived from the classical form of the Hamiltonian. A phase-space formulation of quantum mechanics also extends the Poisson bracket to quantum degrees of freedom, which in our case will be fluctuations of metric components. It is therefore possible to evaluate algebraic relations of the gauge generators, even if they include quantum terms. This procedure results in strict covariance conditions for fluctuating or superposed geometries. The quantum-corrected gauge generators can be viewed as semiclassical approximations of transformations between quantum reference frames in a full field-theory context of space-time.

In this paper, we apply canonical methods of gravity and quantum mechanics in the weak-field regime of static spherically symmetric space-times, reviewed in Sections~\ref{s:SpaceTime} and \ref{s:Quantum}, respectively. By
including phase-space degrees of freedom for quantum fluctuations, we will be able to analyze the geometrical structure and physical properties of the
gravitational field implied by two or more masses in superposition at the same
central point. Our methods are based on \cite{SphSymmMoments}, where the general formalism has been laid out for models of canonical quantum gravity. The previous treatment in a Schwarzschild gauge, relevant for horizon properties, is replaced here by a derivation in isotropic coordinates, which is more suitable for a complete weak-field limit. This choice of various gauges is possible only because we are dealing with a covariant space-time formulation. There is therefore a universal theory of spherically symmetric black holes and their quantum fluctuations that can be used to derive a large set of physical properties. Our new weak-field results, such as novel terms in quantum Newton's potentials, as well as bounds on fluctuations allowed in a covariant context, are contained in Section~\ref{s:Implications}. 

\section{Canonical description of space-time structure}
\label{s:SpaceTime}

The canonical formulation of generally covariant systems describes space-time as an evolving geometry on spacelike hypersurfaces. From the space-time point of view, the spatial geometry on a given hypersurface is determined by the induced spatial metric $q_{ab}$, and its velocity or momentum $p^{ab}$ is related tothe extrinsic curvature of the hypersurface in space-time \cite{DiracHamGR,Katz,ADM}. These relationships define a phase-space structure on the space of geometries. The canonical formulation replaces the derivation of an induced metric and extrinsic curvature from a 4-dimensional line element. It is replaced with coupled evolution equations for two phase-space fields, $q_{ab}$ and $p^{ab}$, together with gauge transformations. These gauge transformations ensure that a 4-dimensional geometry exists, from which these fields may be induced. Geometrically, the gauge transformations correspond to infinitesimal deformations of the spatial hypersurfaces in the normal and tangential directions \cite{Regained}. These deformations are parameterized mathematically by the lapse function and shift vector, as described in this section.

Explicit expressions simplify if spatial geometries are
restricted to be spherically symmetric, which we will assume from now on. The general spatial line element is then given by
\begin{equation}
  {\rm d}s^2= q_{xx}{\rm d}x^2+q_{\vartheta\vartheta}({\rm
    d}\vartheta^2+\sin^2\vartheta{\rm d}\varphi^2)
\end{equation}
with two functions, $q_{xx}$ and $q_{\vartheta\vartheta}$, that depend on the radial coordinate $x$. If a family of hypersurfaces is considered, labeled by time coordinate $t$, then the metric components also depend on $t$.

The definition of the time derivative $\dot{q}_{xx}={\cal L}_t q_{xx}$ as a Lie derivative requires a time direction that relates points on two nearby hypersurfaces with the same spatial coordinates. Such a time direction is not
unique and implies additional free functions. It can be parameterized by the time-evolution vector field
\begin{equation} \label{ta}
  t^a= c(Nn^a+M^a)
\end{equation}
with the speed of light $c$ and the unit normal $n^a$ to a hypersurface. The free components of this space-time vector field are then separated into a normal component, the lapse function $N$, and the three components of a spatial vector field, the shift vector $M^a$, tangential to the hypersurface.

In a covariant theory, the spatial metric, together with the time-evolution vector field, can be used to reconstruct a space-time metric. As inverse metric tensors, the standard relationship is given by $g^{ab}=q^{ab}-n^an^b$, because it implies two properties: (i) $q^{ab}n_a=0$ with the timelike unit normal $n_a$, $n^an_a=-1$, such that the spatial metric is induced on a hypersurface normal to $n^a$. And (ii) $q^{ab}s_a=g^{ab}s_a$ for any vector $s_a$ tangential to the hypersurface, $n^as_a=0$, such that the spatial metric agrees with the space-time metric in this case. Solving (\ref{ta}) for $n^a$ and inserting the result in $g^{ab}$, we obtain
\begin{equation}
  g^{ab}=q^{ab}-\frac{1}{N^2}(t^a/c-M^a)(t^b/c-M^b)\,.
\end{equation}
Inversion of this metric tensor implies the reconstructed space-time line element:
\begin{equation} \label{spacetime}
  {\rm d}s^2= -N^2c^2{\rm d}t^2+ q_{ab}({\rm d}x^a+M^ac{\rm d}t) ({\rm
    d}x^b+M^bc{\rm d}t)\,.
\end{equation}

For (\ref{spacetime}) to provide a generally covariant description of the original canonical theory, we must have slicing independence. Every family of spacelike hypersurfaces (or every time coordinate $t$ that defines spacelike hypersurfaces $t={\rm const}$) implies a time-dependent family of induced spatial metrics and extrinsic curvatures, determined by (\ref{spacetime}). Each such family must evolve in a manner consistent with the canonical equations of motion. Since evolution is canonically determined as Hamilton's equations generated by a Hamilton function on phase space, this function itself must be transformed in a consistent way if the family of hypersurfaces changes. There must, therefore, be a set of Hamiltonian phase-space functions that obey specific algebraic relations through Poisson brackets. The classical theory of general relativity in canonical form can be used to derive these relations, which, in spherical symmetry, turn out to be:
\begin{eqnarray}
 \{D[M_1],D[M_2]\} &=& D[M_1M_2'-M_2 M_1'] \label{DD} \\
 \{H[N],D[M]\} &=& -H[MN'] \label{HD} \\
 \{H[N_1],H[N_2]\} &=& - D[q_{xx}^{-1}(N_1N_2'-N_2 N_1')] \label{HH}
\end{eqnarray}
Here, $H[N]$ is the Hamilton function for a given time component $N$ in (\ref{ta}). $D[M]$ is the generator of a spatial shift tangential to a hypersurface, using the radial component of a spherically symmetric version of
(\ref{ta}).

These relations describe the symmetry of infinitesimal deformations of spatial hypersurfaces. In a physical theory presented in canonical form, abstract generators $D[M]$ and $H[N]$, fulfilling the relations (\ref{DD})--(\ref{HH}), are realized by specific phase-space functions. These functions depend on the geometrical canonical variables $q_{ab}$ and $p^{ab}$, such that their Poisson brackets model (\ref{DD})--(\ref{HH}). Since the underlying symmetries are gauge transformations, the generators must vanish for physical phase-space solutions: $D[M]=0$ for all $M$ and $H[N]=0$ for all $N$. These constraint equations depend on $M$ and $N$, as explicitly shown, but also on $q_{ab}$ and $p^{ab}$ through their phase-space realizations. The constraints do not directly depend on position coordinates $x$ because they are globally defined expressions given by spatial integrals. We use the conventional expression $H[N]$ to denote the Hamiltonian constraint, and similar notation for its quantum-corrected versions. In addition, the notation $H$ by itself  will represent the Hamiltonian constraint's integrand excluding the lapse function. The lapse function always appears as a single factor in the full integrand of $H[N]$ in order to be compatible with relations (\ref{DD})--(\ref{HH}) linear in $D$ and $H$.

Explicit expressions of the gravitational constraints
in spherical symmetry are given by
\begin{equation}
 H[N]=-\frac{c^3}{2G}\int{\rm d}x N(x) \left(\frac{\phi_2p_2^2}{2\sqrt{\phi_1}}+
   2\sqrt{\phi_1} p_1p_2+ \left(1-\left(\frac{\phi_1'}{\phi_2}\right)^2\right)
   \frac{\phi_2}{2\sqrt{\phi_1}}- 2\left(\frac{\phi_1'}{\phi_2}\right)'
   \sqrt{\phi_1}\right)
\end{equation}
and
\begin{equation}
 D[M]= \frac{c^3}{2G}\int{\rm d}xM(x) \left(-\phi_1'p_1+p_2'\phi_2\right)\,.
\end{equation}
We parameterize the metric components as scalar fields
\begin{equation}
  \phi_1=q_{\vartheta\vartheta} \quad\mbox{and}\quad
  \phi_2=2\sqrt{q_{xx}q_{\vartheta\vartheta}}
\end{equation}
with momenta $p_1$ of $\phi_1$ and $p_2$ of $\phi_2$, such that $\{\phi_1(x),p_1(y)\}=2Gc^{-3}\delta(x-y)$ and $\{\phi_2(x),p_2(y)\}=2Gc^{-3}\delta(x-y)$. Here, $G$ is Newton's constant and we use primes to indicate partial derivatives by $x$. We assign units of length to the coordinate $x$ and use unitless angles $\vartheta$ and $\varphi$. Since the line element has units of length squared, $q_{xx}$ is unitless and $q_{\vartheta\vartheta}$ has units of length squared. Therefore, $\phi_1$ has units of length squared as well, while $\phi_2$ has units of length. The momenta's units are then determined by the form of our Poisson brackets: A Poisson bracket of a 1-dimensional field theory has units of $1/\hbar$ (from the product of partial derivatives of position and momentum components) times inverse length from $\delta(x-y)$. Using the expression $\ell_{\rm P}=\sqrt{G\hbar/c^3}$ for the Planck length, the factors of $G/c^3$ in our Poisson brackets therefore provide units of length squared divided by $\hbar$. This is compatible with the units of $\phi_1$ and $\phi_2$ if $p_1$ has units of inverse length and $p_2$ is unitless. Our Hamiltonian constraint then has units of energy divided by $c$. Thus, a Poisson bracket with this expression generates a spatial derivative, rather than a time derivative, as appropriate for a geometrical object.

The relationships between the momenta and time derivatives of the metric components, related to extrinsic curvature, need not be imposed independently. They follow from
Hamilton's equations applied to $\phi_1$ and $\phi_2$: If our time evolution vector field is given by (\ref{ta}) with spherically symmetric components $N$ and $M$, then the time derivatives are
\begin{equation} \label{phi1dot}
  \frac{\dot{\phi}_1}{c} = \{\phi_1,H[N]+D[M]\}= -2N\sqrt{\phi_1} p_2-M\phi_1'
\end{equation}
and
\begin{equation} \label{phi2dot}
  \frac{\dot{\phi}_2}{c}=-N\frac{\phi_2}{\sqrt{\phi_1}}
  p_2-2N\sqrt{\phi_1}p_1-(M\phi_2)'\,.
\end{equation}
These equations can be solved for $p_1$ and $p_2$ in terms of $\dot{\phi}_1$,
$\dot{\phi}_2$, $N$ and $M$, in agreement with expressions for extrinsic curvature. Then the time derivatives of $p_1$ and $p_2$, derived in the same way, determine the evolution equations.

For the weak-field limit, we need the Schwarzschild solution in isotropic coordinates, which can be derived from the canonical equations following the standard Schwarzschild example in \cite{CUP,Foundations}. These coordinates, by definition, imply a line element
\begin{eqnarray}
  {\rm d}s^2&=& -\left(1+\frac{2\Phi(x)}{c^2}+O(\Phi(x)/c^2)\right) c^2{\rm d}t^2\nonumber\\
  &&+ \left(1-\frac{2\Phi(x)}{c^2}+O(\Phi(x)/c^2)\right)\left({\rm
      d}X^2+{\rm d}Y^2+{\rm d}Z^2\right)
\end{eqnarray}
in Cartesian spatial coordinates $(X,Y,Z)$ with $x^2=X^2+Y^2+Z^2$. In this form, the line element directly implies a proper-time interval
\begin{equation}
  {\rm d}\tau=\sqrt{-{\rm d}s^2/c^2}=
  \sqrt{1+2\frac{\Phi(x)}{c^2}-\left(1-2\frac{\Phi(x)}{c^2}\right)
    \frac{|\vec{V}|^2}{c^2}}\;{\rm d}t
\end{equation}
to leading order in Newton's potential, $\Phi(x)/c^2$, and with the velocity vector $\vec{V}={\rm d}\vec{X}/{\rm d}t$ in these coordinates. For small potential and non-relativistic speeds, a further expansion implies
\begin{equation}
  {\rm d}\tau=\left(1+\frac{\Phi (x)}{c^2}-\frac{1}{2}\frac{|\vec{V}|^2}{c^2}\right){\rm d}t
\end{equation}
with a combination of gravitational and special-relativity redshift effects. The metric components in isotropic coordinates therefore directly determine the gravitational potential $\Phi(x)$. (The leading-order potential may be obtained also in standard Schwarzschild coordinates. However, any corrections such as those studied below may require the specific form of isotropic coordinates, in which the gravitational potential and the speed are clearly separated in the non-relativistic limit. In this context, note that Newton's potential is not covariant because it was originally defined assuming an absolute time. In general relativity, this non-covariant concept can be defined only in a specific slicing, which is the one given by isotropic coordinates.)

The isotropic line element can be expressed in the general spherically symmetric form by transforming the original $(x,\vartheta,\varphi)$ in (\ref{spacetime}) to
Cartesian coordinates by
\begin{equation}
  X=x\sin\vartheta\cos\varphi\quad,\quad
  Y=x\sin\vartheta\sin\varphi\quad,\quad Z=x\cos\vartheta\,.
\end{equation}
The line element then takes the form
\begin{eqnarray} \label{isotropic}
  {\rm d}s^2&=& -\left(1+\frac{2\Phi(x)}{c^2}+O(\Phi(x)/c^2)\right) c^2{\rm d}t^2\nonumber\\
      &&+ \left(1-\frac{2\Phi(x)}{c^2}+O(\Phi(x)/c^2)\right)\left({\rm
      d}x^2+x^2({\rm d}\vartheta^2+\sin^2\vartheta{\rm d}\varphi^2)\right)\,.
\end{eqnarray}
By comparing with (\ref{spacetime}), we read off the condition
\begin{equation}
  q_{xx}=\frac{q_{\vartheta\vartheta}}{x^2}
\end{equation}
which can be used to define isotropic coordinates in the canonical formulation, or
\begin{equation} \label{isotropy}
  \phi_2=2\frac{\phi_1}{x}
\end{equation}
for the fields used here. In addition, since the line element in isotropic coordinates is static, we
require that all time derivatives, as well as the ${\rm d}x{\rm d}t$-component $M$ of the line element, vanish. These conditions, taken together, uniquely determine the isotropic slicing into hypersurfaces. In other words, they determine the gauge of the canonical theory.

It immediately follows that the constraint $D[M]=0$, as well as (\ref{phi1dot}) and (\ref{phi2dot}), are identically satisfied. The remaining constraint simplifies to
\begin{equation}
  H[N]_{\rm static}= -\frac{c^3}{2G}\int{\rm d}x N(x)\left(\frac{\sqrt{\phi_1}}{x}+
    \frac{3x(\phi_1')^2}{4\phi_1^{3/2}} -\frac{x\phi_1''}{\sqrt{\phi_1}}
    -\frac{\phi_1'}{\sqrt{\phi_1}} 
    \right)\,.
\end{equation}
This expression vanishes for all $N(x)$ if $\phi_1$ satisfies the differential
equation
\begin{equation} \label{phi12}
  \phi_1''-\frac{3}{4}
  \frac{(\phi_1')^2}{\phi_1}-\frac{\phi_1}{x^2}+\frac{\phi_1'}{x}=0\,. 
\end{equation}
The substitution $f(x)=\sqrt{x}\phi_1(x)^{1/4}$ us employed for the convenience of simplifying this equation to $f''=0$, such that
\begin{equation} \label{phi1}
  \phi_1(x)=\left(a\sqrt{x}+\frac{b}{\sqrt{x}}\right)^4
\end{equation}
with two constants, $a$ and $b$. Therefore,
\begin{equation}
  q_{xx}=\frac{\phi_1}{x^2}= \left(a+\frac{b}{x}\right)^4.
\end{equation}
We can choose $a=1$ without loss of generality because a different value (other than zero) could always be absorbed into the Cartesian coordinates. The weak-field limit then relates
\begin{equation}
  b=\frac{GM}{2c^2}
\end{equation}
to the mass $M$, with Newton's constant $G$. The equation 
\begin{equation} \label{xxSchwarzschild}
    \phi_1(x)= x^2 \left(1+\frac{b}{x}\right)^4=x_{\rm Schwarzschild}^2
\end{equation}
provides the coordinate transformation between isotropic and Schwarzschild coordinates.

The remaining equations to solve are Hamilton's equations for $p_1$ and $p_2$, given by
\begin{eqnarray}
  \frac{\dot{p}_1}{c}&=& -N''\frac{2 \sqrt{\phi _1}}{\phi _2}+N'\frac{2 \phi _1 \phi _2'-\phi _2 \phi _1'}{\sqrt{\phi _1} \phi _2^2}\nonumber\\
&&+N\frac{ -\left(p_2^2+1\right) \phi _2^2+p_1 p_2\phi_1\phi_2+\phi_1 \phi _1' \phi _2'/\phi_2+\frac{1}{4}(\phi_1')^2-\phi_1\phi_1''}{\phi _1^{3/2} \phi _2}
\end{eqnarray}
and
\begin{equation}
  \frac{\dot{p}_2}{c}= \frac{2 N' \sqrt{\phi _1} \phi _1'}{\phi _2^2}+N\frac{ \left(p_2^2+1\right) \phi _2^2-(\phi _1')^2}{2 \sqrt{\phi _1} \phi _2^2}\,.
\end{equation}
With staticity, $p_1=0=\dot{p}_2$, $p_2=0=\dot{p}_2$, and the isotropy condition $\phi_2=2\phi_1/x$, these equations simplify to
\begin{eqnarray}
  0&=&-2x\phi_1N''+ (x  \phi _1'-2 \phi _1 )N'\nonumber\\
&& -\frac{-5 x^2 (\phi _1')^2+4 x \phi _1 \left(\phi _1'+x \phi _1''\right)+4 \phi _1^2}{4 x \phi _1 }N
        \label{p1dot2}
\end{eqnarray}
and
\begin{eqnarray}
  0&=& \frac{ \phi _1^2-\frac{1}{4}x^2 (\phi _1')^2}{\phi _1}N-x^2 \phi _1'N'. \label{p2dot2}
\end{eqnarray}
Using (\ref{phi12}) and (\ref{p2dot2}), equation (\ref{p1dot2}) is turned into
\begin{equation}
  N''=N\left(\frac{(\phi_1')^2}{8\phi_1^2}+ \frac{\phi_1'}{4x\phi_1}-\frac{\phi_1}{x^3\phi_1'}-\frac{1}{2x^2}\right)=\frac{-4b}{(x+b)^2(x-b)}N
\end{equation}
with our solution for $\phi_1(x)$, while (\ref{p2dot2}) takes the form
\begin{equation}
  \frac{N'}{N}=\frac{\phi_1}{x^2\phi_1'}-\frac{\phi_1'}{4\phi_1}= \frac{1}{x-b}-\frac{1}{x+b}\,.
\end{equation}
The functions
\begin{equation}
  N(x)=N_0 \frac{x-b}{x+b}
\end{equation}
solve both equations, with an integration constant $N_0$ that can be absorbed
into the time coordinate. Applying the coordinate transformation
(\ref{xxSchwarzschild}) confirms that this is the correct expression, which
replaces the well-known lapse function in the Schwarzschild line element with
a corresponding expression in isotropic coordinates upon implementing the
coordinate transformation (\ref{xxSchwarzschild}).
Note that the alternative derivation of $N(x)$ is by using the coordinate transformation (\ref{xxSchwarzschild}) in the original Schwarzschild lapse function.

\section{Canonical description of quantum mechanics}
\label{s:Quantum}
Given a quantum system with canonically conjugate basic operators $\hat{q}$ and $\hat{p}$ with $[\hat{q},\hat{p}]=i\hbar$, it is possible to equip states and operators with geometrical meaning on a quantum phase space. We first interpret operators $\hat{O}$ as functions $f_{\hat{O}}$ on the space of states $\psi$, given by the evaluation $f_{\hat{O}}(\psi)=\langle\psi,\hat{O}\psi\rangle$. There is a unique complex-valued function with a dense domain of $\psi$ for every operator $\hat{O}$ acting on the Hilbert space of the system. In general, this function is not defined for all $\psi$ because expectation values of unbounded operators may be infinite in some states, but it is defined for a dense subset in the topological sense. Useful examples for the following developments are the functions
\begin{eqnarray}
  q&=&f_{\hat{q}}=\langle\hat{q}\rangle\\
  p&=&f_{\hat{p}}=\langle\hat{p}\rangle
\end{eqnarray}
as well as functions for products of multiple $\hat{q}$'s and $\hat{p}$'s, from which we can construct the central moments
\begin{equation} \label{Delta}
 \Delta(q^np^m)=\langle(\hat{q}-\langle\hat{q}\rangle)^n
(\hat{p}-\langle\hat{p}\rangle)^m\rangle_{\rm symm}
\end{equation}
with operator products in completely symmetric (or Weyl) ordering. 

\subsection{Dynamics}

On this function space, we define a Poisson bracket by
\begin{equation} 
  \{f_{\hat{A}},f_{\hat{B}}\}=\frac{f_{[\hat{A},\hat{B}]}}{i\hbar}
\end{equation}
or, equivalently,
\begin{equation}
  \{\langle\hat{A}\rangle,\langle\hat{B}\rangle\}=
 \frac{\langle[\hat{A},\hat{B}]\rangle}{i\hbar}
\end{equation}
which is extended to products of functions by using the Leibniz rule. This Poisson bracket turns Ehrenfest's theorem into a statement about Hamiltonian dynamics: The evolution of expectation-value functions is given by Hamilton's equations generated by the expectation value
\begin{equation}
  H_{\rm eff}= f_{\hat{H}}=\langle\hat{H}\rangle
\end{equation}
because
\begin{equation} \label{Poisson}
 \{\langle\hat{A}\rangle,H_{\rm eff}\}=
 \frac{\langle[\hat{A},\hat{H}]\rangle}{i\hbar} = \frac{{\rm
     d}\langle\hat{A}\rangle}{{\rm d}t}\,.
\end{equation}

If we use the basic expectation values $q$ and $p$, together with central moments, as coordinates on the quantum phase space, we should write the effective Hamiltonian $H_{\rm eff}$ as a function of these variables. We can do so by using a formal Taylor expansion in
\begin{eqnarray}
 H_{\rm eff} &=& \langle H(\hat{q},\hat{p})\rangle= \langle
 H(q+(\hat{q}-q),p+(\hat{p}-p)\rangle\nonumber\\
 &=& H(q,p)+\sum_{n+m=2}^{N} \frac{1}{n!m!}
 \frac{\partial^{n+m}H(q,p)}{\partial q^n\partial p^m}
 \Delta(q^np^m) \label{Heffgen} 
\end{eqnarray}
assuming that the Hamilton operator is Weyl ordered. (If this is not the case, there will be additional reordering terms that explicitly depend on $\hbar$.)

The Poisson bracket of two moments is not constant, as can be seen by a direct calculation \cite{EffAc,HigherMoments}. Therefore, the moments are not canonical coordinates on the quantum phase space. The Darboux theorem and its generalization to Poisson manifolds \cite{Arnold,Weinstein} guarantees the existence of local canonical coordinates with the standard values zero and one of basic Poisson brackets. However, suitable coordinates of this form may be difficult to derive in general. For low orders, several examples exist. We are particularly interested in the second-order canonical parameterization \cite{VariationalEffAc,EnvQuantumChaos,QHDTunneling,CQC}
\begin{equation} \label{sps}
 \Delta(q^2) = s^2 \quad,\quad \Delta(qp)=sp_s \quad,\quad \Delta(p^2)=
 p_s^2+\frac{U}{s^2}
\end{equation}
where $s$ and $p_s$ are canonically conjugate, $\{s,p_s\}=1$, and $U$ has vanishing Poisson brackets with both $s$ and $p_s$. The uncertainty principle restricts the values of $U$ by $U\geq \hbar^2/4$. Further examples have been derived in \cite{Bosonize,EffPotRealize} for moments of up to fourth order or for two independent degrees of freedom. With these canonical variables, the effective Hamiltonian of a 1-dimensional mechanical system takes the form
\begin{equation}
 H_{\rm eff}= \frac{p^2}{2m}+ \frac{p_s^2}{2m}+ \frac{U}{2ms^2}+ V(q)+
 \frac{1}{2} V''(q)s^2\,,
\end{equation}
which can directly be used to compute Hamiltonian equations of motion, with coupling terms (or quantum back-reaction) between $(q,p)$ and $(s,p_s)$. 

The example of the harmonic oscillator, where $V(q)=\frac{1}{2}m\omega^2q^2$, shows that $U$ determines the value of zero-point energies: In this case, the Hamiltonian contains the $s$-dependent potential \begin{equation}\label{Vs}
    V_s(s)= \frac{U}{2ms^2}+\frac{1}{2}m\omega^2s^2
\end{equation}
which is minimized by 
\begin{equation}
s^2=\frac{\sqrt{U}}{m\omega}=\frac{\hbar}{2m\omega}\,,
\end{equation}
using the minimal value for $U$ in the second step. At this value of $s$, the effective potential evaluates to a constant contribution $V_s(s)=\omega\sqrt{U}=\frac{1}{2}\hbar\omega$. We will consider this role of $U$ in our extension to a 1-dimensional field theory in the radial direction of our spherically symmetric models. Quantization then require a suitable regularization procedure that subtracts zero-point energies or vacuum expectation values from the Hamiltonian, just as in standard quantum field theory.

For relativistic systems and the physics of space-time, we need two extensions of this formalism: a canonical formulation of constraint operators such as $\hat{H}$ and $\hat{D}$, and a suitable description for quantum field
theories. The first extension \cite{EffCons,EffConsRel,EffConsQBR} is technically involved because it requires general properties of constrained systems, but it follows directly from a basic definition of an effective constraint
\begin{equation}
  C_{\rm eff}=f_{\hat{C}}=\langle\hat{C}\rangle
\end{equation}
for every constraint operator $\hat{C}$ in the system. Physical states then define a submanifold of the quantum phase space defined by $C_{\rm eff}=0$ for each constraint. 

However, it turns out that this restriction is not sufficient because the quantum constraint equation $\hat{C}|\psi\rangle=0$ implies that not only does
$\langle\psi,\hat{C}\psi\rangle$ vanish, but also $\langle\psi, \hat{O}\hat{C}\psi\rangle$ for any operator $\hat{O}$. A single quantum constraint therefore implies infinitely many constrained expectation values. If we parameterize these constraints by
\begin{equation} \label{HOC}
  \langle (\hat{q}-q)^n(\hat{p}-p)^m\hat{C}\rangle=0
\end{equation}
with positive integers $n$ and $m$, there is a finite number of such constraints for any set of central moments up to some finite order. Such a system of constraints can be treated by the general methods for constrained systems as given in \cite{DirQuant}. It constitutes a semiclassical approximation of infinitesimal transformations between quantum reference frames of space-time.

In spherically symmetric gravitational systems, we have two constraints, $H[N]$ and $D[M]$. For static solutions, $D$ and its higher-order versions (\ref{HOC}) vanish identically, and higher-order versions of $H$ need only be
considered for the $\phi$-fields, but not for their momenta, such that $m=0$ in (\ref{HOC}). For constraints on second-order moments, including quantum fluctuations, it is sufficient to consider only $n=0$ and $n=1$ in (\ref{HOC}). Moreover, the isotropy condition strictly relates $\phi_1$ and $\phi_2$, and therefore their fluctuations.

\subsection{Ingredients from quantum field theory}

In order to derive the relevant constraints and higher-order versions, we need to determine how quantum parameters such as $s$ and $p_s$ can be implemented for fields. In general, quantum fields may be in a state with non-zero
correlations between their values of different positions, which would require infinitely many correlation fields. Here, for a first analysis that aims mainly to include quantum fluctuations, it will be sufficient to ignore such non-local correlations and only implement a version of $(s,p_s)$ at each point on the radial line. Therefore, we will introduce an additional fluctuation field, $\phi_3$ with a canonical momentum $p_3$, such that
\begin{equation} \label{phip}
 \Delta(\phi_2^2)=\phi_3^2\quad,\quad \Delta(\phi_2p_2)=\phi_3p_3 \quad,\quad 
 \Delta(p_2^2)= p_3^2+\frac{U}{\phi_3^2}
\end{equation}
independently at each radial position $x$. The treatment of $U$ now requires some care because, as shown by our discussion of the effective potential (\ref{Vs}) of the harmonic oscillator, this parameter is related to zero-point energies. For a quantum field theory on Minkowski space-time, we would simply remove the $U$-term from any Hamiltonian as a subtraction of zero-point energies. In our background-independent treatment, however, we can only assume that we will be near Minkowski space-time in a suitable asymptotic region of small curvature, close to the regime where we will apply our main weak-field analysis. In regions of curved space-time that are not strictly Minkowskian, which are relevant for any radial dependence of quantities such an effective gravitational potential, the vacuum state and corresponding zero-point energies are not uniquely defined. As a result, we cannot simply remove the entire $U$-term from our Hamiltonian but rather subtract only the Minkowski limit, as we will see in more detail when we discuss the specific Hamiltonian.

For now, we conclude that some $U$-dependent contributions are likely to remain in the Hamiltonian and are fully subtracted only in the asymptotic Minkowski region at $x\to\infty$. The presence of a subtraction implies that we should no longer subject $U$ to the inequality $U\geq\hbar^2/4$ from quantum mechanics. We will therefore impose only positivity of $U$ without using a general non-zero lower limit. And since the subtraction is complete only at $x\to\infty$, we expect that remaining effects described by $U$ may be given by a function $U(x)$ of $x$ that should be determined by consistency conditions within the quantum theory but cannot be known a priori. 

Another basic implication of a treatment of our model as a quantum field theory is that quantizations of $\phi_2(x)$ and $p_2(x)$ are operator-valued distributions. Their commutator equals $i\hbar G/c^3$ times a 1-dimensional delta function, where the factor of $G/c^3$ comes from our Poisson bracket.  Combining all factors, including an inverse length scale for units of the delta function, the commutator has units determined by $G\hbar/(c^3L)=\ell_{\rm P}^2/L$ with the Planck length $\ell_{\rm P}$ and a relevant macroscopic, $\hbar$-independent length $L$ such as the Schwarzschild radius. The uncertainty relation, derived from this field-theory commutator, then implies that $U=\Delta(\phi_2^2)(\Delta(p_2^2)-p_3^2)$ has units of $\ell_{\rm P}^4/L^2$. This result differs from the units of $\hbar^2/4$ obtained for $U$ in quantum mechanics. Therefore, just based on dimensional reasoning, it is now impossible to impose a lower bound of $\hbar^2/4$ on $U$ in a quantum field theory. This observation provides further motivation to require only positivity of $U(x)$ but not a stronger non-zero lower bound.

Following our discussion of units for the classical fields, the individual fields used here have units of length for $\phi_2$ and $\phi_3$, inherited from the metric, and no units for $p_2$ and $p_3$. The quantum fields $\phi_3$ and $p_3$, defined as square roots of second-order moments of a state, are expected to be proportional to $\sqrt{\hbar}$ for semiclassical solutions. These conditions are consistent with one another if $\phi_3\sim \ell_{\rm P}$ and $p_3\sim\ell_{\rm P}/L$. For $U$, we then obtain $U\sim \ell_{\rm P}^4/L^2$, in agreement with the result suggested by field commutators.

\subsection{Gauge choice and constraints}

Finally, we must implement the isotropy condition (\ref{isotropy}). Since
momentum fluctuations may contribute non-zero terms such as $U/\phi_3^2$, even in static situations, we first find the accompanying equations to $2\phi_1-x\phi_2=0$, a corresponding condition on momenta. For the isotropy condition to be preserved over time and in the zero $M$ limit, the equations of motion (\ref{phi1dot}) and (\ref{phi2dot}) imply that the momenta must be related by
\begin{equation}\label{isotropyp}
  xp_1-p_2=0\,.
\end{equation}
This pair of conditions is second class: The smeared Poisson bracket
\begin{equation}
  \{2\phi_1(x)-x\phi_2(x),\int \lambda(y)(yp_1(y)-p_2(y)){\rm d}y\}= 6\frac{G}{c^3}x\lambda(x)
\end{equation}
is non-zero for $x\not=0$. We should therefore impose the two conditions contained in the second-class pair before we quantize, for instance by eliminating $(\phi_1,p_1)$ in favor of $(\phi_2,p_2)$. The remaining field will then be supplied by quantum variables, as in (\ref{phip}).
(After imposing the second-class constraints, the Dirac bracket of $\phi_2$ and $p_2$ is $2/3$ times the Poisson bracket. The additional factor provides a constant rescaling of all time derivatives, which we may ignore since we are interested in static configurations.)

The reduced classical Hamiltonian constraint after imposing the second-class isotropy conditions equals 
\begin{equation}
  H[N]=-\frac{c^3}{2G}\int{\rm d}x N(x) \sqrt{\frac{\phi_2}{2x}} \left( 3 p_2^2+
    \frac{3}{4} -\frac{3}{2} \frac{x\phi_2'}{\phi_2}+
    \frac{3}{4}\frac{ x^2(\phi_2')^2}{\phi_2^2}
    -  \frac{x^2\phi_2''}{\phi_2}\right)
\end{equation}
A Taylor expansion as in (\ref{Heffgen}) then leads to
\begin{equation}
 \bar{H}[N]= H[N]+H_2[N]
\end{equation}
with the classical $H[N]=\int{\rm d}x NH$ and and a quantum correction
\begin{eqnarray}
 H_2[N] &=& \int{\rm d}x N(x)
 \left(\frac{1}{2}\frac{\partial^2H}{\partial p_2^2}
   \left(p_3^2+\frac{U}{\phi_3^2}\right)-H_0+
   \frac{\partial^2H}{\partial\phi_2\partial p_2}\phi_3p_3
\right.\nonumber\\
  &&\qquad\qquad\left.+\frac{1}{2}\frac{\partial^2H}{\partial\phi_2^2} \phi_3^2+
   \frac{\partial^2H}{\partial\phi_2\partial\phi_2'}
            \phi_3\phi_3'+ \frac{1}{2}\frac{\partial^2H}{(\partial\phi_2')^2} (\phi_3')^2+
   \frac{\partial^2H}{\partial\phi_2\partial\phi_2''}
            \phi_3\phi_3''\right)\,.
\end{eqnarray}
Partial derivatives of $H$ follow the form of the expanded effective Hamiltonian in (\ref{Heffgen}), and $H_0$ is the vacuum expectation value to be subtracted from the Hamiltonian. We can write this terms as
\begin{equation}
    H_0=\frac{1}{2}\lim_{b\to0} \frac{U}{\phi_3^2}\frac{\partial^2H}{\partial p_2^2} 
\end{equation}
if we use $b$, related to the black-hole mass in specific solutions, as a parameter that determines deviations from Minkowski space-time. Partial derivatives of the classical Hamiltonian then implies the following coefficients:
\begin{eqnarray}
H_2[N]&=& -\frac{c^3}{2G}\int{\rm d}xN(x) \left(3\frac{\sqrt{\phi_2}p_3^2}{\sqrt{2x}}+
   3\frac{\phi_3p_2p_3}{\sqrt{2x\phi_2}}-
    \frac{3}{32}\frac{\phi_3^2}{\sqrt{2x}\phi_2^{3/2}}-
    \frac{9}{16}\frac{\sqrt{x}}{\sqrt{2}\phi_2^{5/2}} \phi_2'\phi_3^2\right.\nonumber\\
 &&  \qquad\qquad + \frac{45}{32} \frac{x^{3/2}}{\sqrt{2}\phi_2^{7/2}} (\phi_2')^2\phi_3^2
    -\frac{3}{8}\frac{x^{3/2}}{\sqrt{2}\phi_2^{5/2}}\phi_2''\phi_3^2
    +\frac{3}{4}\frac{\sqrt{x}}{\sqrt{2}\phi_2^{3/2}} \phi_3\phi_3'
  \label{H2}\\
&&\qquad\qquad
    -   \frac{9}{4} \frac{x^{3/2}}{\sqrt{2}\phi_2^{5/2}}\phi_2'\phi_3\phi_3'+
   +\frac{3}{4} \sqrt{\frac{\phi_2}{2x}} \frac{x^2}{\phi_2^2}(\phi_3')^2+
   \frac{1}{2}\frac{x^{3/2}}{\sqrt{2}\phi_2^{3/2}} \phi_3\phi_3''\nonumber\\
   &&\qquad\qquad\left. +3\left(\sqrt{\frac{\phi_2}{2x}} \frac{U}{\phi_3^2}-\lim_{b\to0}\frac{U}{\phi_3^2}\right) \right)\,.\nonumber 
\end{eqnarray}
In the vacuum subtraction, we used the Minkowski limit $\phi_2\to 2x$ of $\phi_2$. At this point, the $b$-dependence of $\phi_3$ and possibly $U$ remains to be determined from field equation.

The only higher-order constraint we need to consider after the reduction is given by
\begin{equation}
 H_{\phi_2}[L]=\langle(\hat{\phi}_2-\langle\hat{\phi}_2\rangle)
 \hat{H}[L]\rangle\,.  
\end{equation}
Also here, a Taylor expansion provides the constraint as a function of moments:
\begin{eqnarray}
 &&H_{\phi_2}[L] = \int{\rm d}x L(x) \left(\frac{\partial H}{\partial\phi_2}
   \phi_3^2+ \frac{\partial H}{\partial \phi_2'} \phi_3\phi_3'+ \frac{\partial
                   H}{\partial \phi_2''} \phi_3\phi_3''+ \frac{\partial 
     H}{\partial p_2} \phi_3p_3\right)\nonumber\\
&=&-\frac{c^3}{2G}\int{\rm d}x L(x)\Biggl( \left(\frac{3}{2}\frac{p_2^2}{\sqrt{2x\phi_2}}+
    \frac{39}{8} \frac{1}{\sqrt{2x\phi_2}}+\frac{3}{4} \frac{\sqrt{x}}{\sqrt{2}\phi_2^{3/2}}\phi_2'-
    \frac{9}{8}\frac{x^{3/2}}{\sqrt{2}\phi_2^{5/2}} (\phi_2')^2
    + \frac{x^{3/2}}{2\sqrt{2}\phi_2^{3/2}}\phi_2'' \right) \phi_3^2\nonumber\\
&& +\left( - \frac{6}{4}\sqrt{\frac{x}{2\phi_2}}+
    \frac{6}{4}\frac{x^{3/2}}{\sqrt{2}\phi_2^{3/2}} \phi_2' \right)
   \phi_3\phi_3' 
    -  \frac{x^{3/2}}{\sqrt{2\phi_2}}\phi_3\phi_3''
   + 5\sqrt{\frac{2\phi_2}{x}} p_2\phi_3p_3\Biggr) \,. \label{Hphi2}
\end{eqnarray}

\subsection{Solutions}

Inserting the background solution
\begin{equation}
  \sqrt{\frac{\phi_2}{2x}}= \left(1+\frac{b}{x}\right)^2
\end{equation}
based on (\ref{phi1}) with $a=1$ and (\ref{isotropy}),
as well as $p_2=0$, we have
\begin{eqnarray} \label{Hphi2x}
 &&H_{\phi_2}[L]\\
  &=& - \frac{c^3}{2G}\int{\rm d}x  \frac{L\phi_3}{x(1+b/x)^4}\Biggl(3 \frac{b}{x}\left(1-b/x\right) \phi _3
  -3  \frac{b}{x}\left(1+b/x\right)  x \phi _3'
  -\frac{1}{2} \left(1+b/x\right)^2   x^2\phi _3''\Biggr)\,. \nonumber
\end{eqnarray}
The constraint $H_{\phi_2}[L]=0$ is fulfilled  for all $L(x)$ if $\phi_3$ obeys the second-order differential equation
\begin{equation} \label{phi3exp}
    3 \frac{b}{x}\left(1-b/x\right) \phi _3
  -3  \frac{b}{x}\left(1+b/x\right)  x \phi _3'
  -\frac{1}{2} \left(1+b/x\right)^2   x^2\phi _3''=0
\end{equation}
which does not seem to have simple closed-form solutions. However, thanks to the factor of $b/x$ in each of the two lower-order derivative terms, there is a unique solution (up to a constant multiplicative factor) which permits an asymptotic expansion $\phi_3(x)\propto 1+a_1/x+a_2/x^2+\cdots$, compatible with the weak-field limit. The coefficients $a_1$, $a_2$, \ldots can be computed iteratively from
\begin{equation}
    (3b-a_1)x^{-1}+((3ba_1-3b^2)+3ba_1-(2a_1b+3a_2))x^{-2}+O(x^{-3})=0
\end{equation}
and collecting terms with the same factor of $x^{-n}$. The three individual terms in the second parenthesis result from the three independent derivative expressions in (\ref{phi3exp}). From the first two terms of the expansion, we obtain the solution
\begin{equation}
    \phi_3(x)=C\left(1+\frac{3b}{x}+ \frac{3b^2}{x^2}+\cdots\right)\,.
\end{equation}
To first order in $b/x$, this result is consistent with the corresponding solution derived for Schwarzschild coordinates in \cite{SphSymmMoments}. Since $\phi_3$ should have units of length, like $\phi_2$, and be proportional to $\sqrt{\hbar}$ for semiclassical solutions, we have $C\propto \ell_{\rm P}$.

We recall that the new field $\phi_3$ describes quantum fluctuations of the metric component $\phi_2$, or of the full spatial metric, since $\phi_1$ is strictly related to $\phi_2$ by the isotropy condition (\ref{isotropy}).  In order to derive implications of quantum fluctuations on Newton's potential, we have to solve the remaining constraint and evolution equations for a correction $\delta N$ of the classical lapse function, which happens to depend on the correction $\delta\phi_2$ of $\phi_2$ implied by quantum effects. The required equations are considerably longer than those used for our derivation of $\phi_3$ and are therefore collected in Appendix~\ref{app}. 
For the weak-field behavior, it is sufficient to derive the asymptotic form of solutions for $\phi_2$ and $N$ for large $x$. Given the leading orders $\phi_2\sim x$ and $N\sim 1$, our equations can be solved for the coefficients $c_i$ and $d_j$ in 
\begin{equation}
    \phi_2=\phi_2^{(0)}+ c_1+\frac{c_2}{x}+\cdots\quad,\quad N=N^{(0)}+\frac{d_1}{x}+\frac{d_2}{x^2}+\cdots
\end{equation}
where $\phi_2^{(0)}$ and $N^{(0)}$ are the classical solutions in isotropic coordinates. Corrections to Newton's potential can be read off directly from $N(x)^2=1+2\Phi(x)/c^2$, in which $\Phi(x)$ is the quantum-corrected gravitational potential. 

A general implication of the dynamical equations, in particular using $\dot{p}_2=0$ for static solutions and to leading order in $1/x$, is that $d_1=0$; see equation~(\ref{p2dotExpand}) in the appendix. The leading term in Newton's potential is therefore unmodified, but there are higher-order terms in $1/x$. The remaining coefficients are determined by terms in $\dot{p}_2=0$ of higher order in $1/x$, coupled to the Hamiltonian constraint. With the quantum extension, the latter depends on the function $U$, for which we should assume a power-law form $U(x)\propto x^{-\kappa}$, with some $\kappa\geq 0$, in order for a weak-field expansion to exist. It turns out that there are also rather strong reality conditions implied by the explicit terms of the Hamiltonian constraint, which is quadratic in $\phi_3$ and the perturbation $\phi_2-\phi_2^{(0)}$ and should be compatible with real $c_i$. A direct evaluation shows that these conditions can be realized only if $U(x)=U_0/x$ with a constant $U_0$, but not for constant $U$ or $U\propto 1/x^2$. Relevant conditions can be seen, for instance, in (\ref{HExpand}), derived for $U(x)=U_0/x$: The second-order term in this expansion of $H[N]$ in $1/x$ has mixed signs, such that there is a range of quantum parameters for which $c_1$ is real. For $U(x)=U$ or $U(x)=U_0/x^{\kappa}$ with $\kappa\geq2$, by contrast, all signs in this term would be the same, implying imaginary solutions for $c_1$. These properties are discussed in more detail in App.~\ref{s:Asymp}.

We evaluate the expansions (\ref{HExpand}) and (\ref{p2dotExpand}) for the coefficients $c_i$ and $d_i$ up to $i=3$, or third order in $1/x$, using the power-law $U(x)=U_0/x$ that allows real solutions.
The new constant $U_0$ introduced by this parameterization should, according to the discussion following equation (\ref{phip}), have units of $\hbar^2/L$ with a classical length scale $L$. A natural choice for the latter is the Schwarzschild radius or  $b$ in the present context. The same classical parameter also appears in the equations for $c_i$ and $d_i$ because they have been obtained after perturbing $\phi_2$ and $N$ around their $b$-dependent classical values. 
The coupled set of conditions for $d_2$ and $d_3$ in $N$ as well as
$c_1$, $c_2$ and $c_3$ 
in $\phi_2$, contains linear and quadratic equations which are solved by
\begin{eqnarray}
    c_1&=& \epsilon \frac{8\sqrt{bU_0}}{C} \sqrt{1-C^4/(64bU_0)} \label{c1}\\
    c_2&=& \epsilon  b\frac{3\sqrt{bU_0}}{8C} \frac{70-9C^4/(8bU_0)}{\sqrt{1-C^4/(64bU_0)}}\\
    d_2&=& \epsilon  b\frac{\sqrt{bU_0}}{96C} \frac{394- 47C^4/(8bU_0)}{\sqrt{1-C^4/(64bU_0)}}\label{d2}\\
    c_3&=& \epsilon b^2 \frac{\sqrt{bU_0}}{80C}\frac{2445 -79 C^4/(b U_0)+163 C^8/(256b^2U_0^2)}{ \left(1-C^4/(64bU_0)\right){}^{3/2}}\label{c3}\\
    d_3&=& -\epsilon b^2 \frac{\sqrt{bU_0}}{640C} \frac{9735 -589 C^4 /(2bU_0)+569 C^8/(256b^2U_0^2)}{ \left(1-C^4/(64bU_0)\right){}^{3/2}}\label{d3}\,.
\end{eqnarray}
There is a single sign choice $\epsilon=\pm1$ in $c_1$, which is not fixed by the constraint but determines the sign choices in $c_2$, $c_3$, $d_2$, and $d_3$.

\section{Implications}
\label{s:Implications}
The values (\ref{c1})--(\ref{d2}) have several unexpected implications. The solution for $d_2$ immediately gives the leading quantum correction to Newton's potential and can therefore be compared (although not directly, as we will see) with calculations in perturbative quantum gravity. The same coefficient, together with $c_1$ and $c_2$, implies characteristic features of a line element that may be used for a covariant description of superpositions of central masses.

\subsection{Effective potentials}

First, using $d_1=0$, the quantum-corrected Newton potential is given by
\begin{equation} \label{V}
    V(x)=\frac{c^2}{2}(N(x)^2-1)\approx-\frac{GM}{x}+\frac{G^2M^2}{x^2c^2} +\frac{d_2}{2x^2}+O(x^{-3})
\end{equation}
where we used $b=GM/(2c^2)$. We also expanded the lapse function $N^{(0)}(x)$ in isotropic coordinates to second order in $x^{-1}$. Since $d_1=0$, Newton's constant is not renormalized.  The leading quantum correction $d_2$ from (\ref{d2}) is of the order $GM\ell_{\rm P}/c^2$ because $C$, for a semiclassical state, is of the order $\sqrt{\hbar}\propto \ell_{\rm P}$ and $\sqrt{bU_0}$ is of the order $C^2$. 
The classical term in (\ref{V}) quadratic in $x^{-1}$ agrees with the perturbative result from \cite{EffectiveNewton}, and it has the same origin. However, the leading quantum correction in this case is of the order $M\ell_{\rm P}^2/x^3$, which is smaller than our $M\ell_{\rm P}/x^2$ for generic quantum corrections. If we extend $\delta\phi_2$ and $\delta N$ to higher orders in $x^{-1}$, we obtain additional terms of the order $b^n\ell_{\rm P}$ with integers $n>1$. Their dependence on $\ell_{\rm P}$ or $\hbar$ remains of first order, which is fixed in our approximation by using second-order moments of order $\hbar$ in the constraint. As a quadratic expression in $\delta\phi_2$ and $\delta N$, the constraint then implies corrections in $\delta N$ of the order $\sqrt{\hbar}$. Higher orders in $\sqrt{\hbar}$ would require higher moments.

We could try to impose $d_2=0$, eliminating our larger term compared with perturbative quantum field theory, by choosing suitable values for $C$ and $U_0$. This is possible only if
\begin{equation}\label{CU}
 C^4=\frac{3152bU_0}{47}\approx 67.1 bU_0\,,
\end{equation}
but this value is not compatible with the upper bound on $C^4<64bU_0$ implied by the square root in the denominator of $d_2$ being real. Therefore, our quantum corrections are always greater than what is expected from perturbative quantum gravity as derived (with varying results) for instance in \cite{EffectiveNewton,EffectiveGR,LongRangeQG,QGNewton,NewtonOneLoop,TwoMasses,NewtonKerr,EffNewtonCoulomb}. This discrepancy can be explained by the fact that we are considering different physical settings, such that the underlying states, which determine quantum corrections, need not be the same. Here, we use a generic semiclassical state parameterized by its moments. The moments, derived from the gravitational constraints, describe an entire black hole (or a superposition of black holes) that has been formed by some collapse process and eventually settled down to a static configuration. In contrast, the result from perturbative quantum gravity implicitly refers to 2-particle states close to the interacting vacuum. This vacuum is itself close to the Gaussian vacuum state of a free field theory in perturbative situations.  Moreover, perturbative calculations are usually done for two masses on Minkowski space-time, while our results are for the effective Newton potential experienced by a light test mass in the curved background of a heavy mass. The physical settings are therefore different, which means that differing results are not problematic. Coefficients, and even orders of quantum corrections, may well be different in our case.

We already referred to an upper bound on $C$, implied by reality conditions on the expansion coefficients (\ref{c1})--(\ref{d2}). This upper bound is of interest and is explicitly given by
\begin{equation} \label{Ineq}
    \phi_3\sim C\leq 2\sqrt{2} (bU_0)^{1/4}\,.
\end{equation}
If the classical length scale $L$ in $U_0$ is equated with $b$, we have $(bU_0)^{1/4}\sim \sqrt{\hbar}$, which is the expected order of $C$ for semiclassical solutions. With the required length units for $\phi_2$ and $\phi_3$, we can parameterize $bU_0$ as $u\ell_{\rm P}^4$ with a unitless number $u$. Given the inequality in (\ref{Ineq}), a space-time geometry is compatible with such quantum corrections only if $C$, or the quantum fluctuations $\phi_3$ of $\phi_2$, are subject to the upper bound 
\begin{equation} \label{Ineqp}
    \Delta\phi_2\leq 2\sqrt{2}u^{1/4}\ell_{\rm P}\,.
\end{equation}
Since  (\ref{phip}) implies
\begin{equation}
    (\Delta p_2)^2(\Delta\phi_2)^2=U+p_3^2\phi_3^2\geq U=\frac{u\ell_{\rm P}^4}{b^2}\,,
\end{equation}
momentum fluctuations are bounded from below by
\begin{equation}
    \Delta p_2\geq \frac{\sqrt{u}\ell_{\rm P}^2}{b\Delta\phi_2}\geq \frac{u^{1/4}\ell_{\rm P}}{2\sqrt{2}b}\,.
\end{equation}
Static solutions as derived here, which by assumption have a vanishing expectation value of $p_2$, can therefore exist only with non-zero momentum fluctuations. The geometrical meaning of the momentum as extrinsic curvature of spatial slices suggests that quantum black holes that are static on average are, in fact, superpositions of collapsing and expanding geometries, corresponding to wave functions that have support on positive and negative values of $p_2$. An open question is whether such quantum oscillations would imply the emission of gravitational waves if non-spherical perturbations are included in our models.  Since $u\propto b^{1/4}$, according to its definition, the lower bound on $\Delta p_2$ decreases with the mass and is negligible for Schwarzschild radii much greater than the Planck length. This superposition effect in the momentum and possible quantum oscillations would therefore be relevant only for microscopic or primordial black holes.

Compared with the previous \cite{SphSymmMoments}, on which the present paper is based, we have performed calculations in isotropic rather than Schwarzschild coordinates. This change of coordinates implies that a clear analysis of the weak-field limit could be performed with an unambiguous definition of Newton's potential. Crucially, our space-time description made it possible to consider different coordinate systems within the same model, which is a requirement for a simultaneous interpretation of physical effects in the weak-field regime (derived here) as well as close to the horizon of a black hole (considered in \cite{SphSymmMoments}). Our novel weak-field results qualitatively confirm the fall-off behavior of the function $U(x)$ which previously was found only numerically. They also led us to re-interpret this function by introducing a suitable length scale, $L$, implied by the density-behavior of fields compared with point particles in quantum mechanics, as discussed in detail in the passage following equation (\ref{phip}). The function $U(x)$ here therefore has units of $\hbar^2/L$, rather than $\hbar^2$, which is important for estimating orders of magnitude for quantum corrections. 

\subsection{Black-hole superpositions}
We have obtained specific expressions (\ref{c1})--(\ref{d2}), together with $d_1=0$, for the expansion coefficients $c_i$ of $\delta\phi_2$ and $d_i$ of $\delta N$. In canonical gravity, $\phi_2$ is a phase-space degree of freedom that would be quantized in canonical quantum gravity and therefore have quantum fluctuations. Here, these fluctuations are described by a new independent field $\phi_3$ and determined in our solutions by the parameter $C$. Solving the Hamiltonian constraint also implies a dependence of $c_i$ and $d_i$ on the spatial function $U(x)$ that can be interpreted as the uncertainty product, imposing a lower bound $(\Delta\phi_2)^2(\Delta p_2)^2-\phi_3^2p_3^2$ as an expression of the uncertainty relation. 

In contrast to the $c_i$, the coefficients $d_i$ in an expansion of $\delta N$ do not appear in a phase-space degree of freedom because $N$ does not play such a role in the formulation of canonical gravity used here. (An extended phase-space formulation could include $N$ as a phase-space degree of freedom, but only in a limited role because its momentum would be constrained to vanish.) Canonically, the coefficients $c_i$ and $d_i$, respectively, therefore play different roles in how they may be related to wave functions or quantum fluctuations. The $c_i$ in $\delta\phi_2$ have a more direct relationship with quantum fluctuations than the $d_i$ in $\delta N$. In the derivation of Newton's potential, however, the $d_i$ are more relevant because the gravitational potential in the weak-field limit appears in $N$ to leading order. 

An effective Newton's potential therefore depends on quantum fluctuations in a rather indirect way. One such dependence is given by the relationships between $c_i$ and $d_i$ implied by the constraints and staticity conditions, which we solved in order to obtain our solutions. Moreover, if we use the full space-time metric, all the coefficients $c_1$, $c_2$, $d_1$ and $d_2$ introduced in our solution procedure appear in the corrected metric components through $\delta N$ and $\delta\phi_2$ in isotropic coordinates. The corresponding line element is of the form
\begin{eqnarray}
    {\rm d}s^2&=& - \left(N^{(0)}+ \frac{d_1}{x}+\frac{d_2}{x^2}+O(x^{-3})\right)^2 c^2{\rm d}t^2\nonumber\\
    &&+ \left(\frac{\phi_2^{(0)}}{2x}+\frac{c_1}{2x}+\frac{c_2}{2x^2}+O(x^{-3})\right) \left({\rm d}x^2+x^2({\rm d}\vartheta^2+\sin^2\vartheta{\rm d}\varphi^2)\right)\,.
\end{eqnarray}
One implication is that the coefficients $c_i$ in $\delta\phi_2$ change the radial length according to the underlying geometry. If a corrected Newton's potential is expressed as a function of the geometrical radius $r=\int (\phi_2^{(0)}+\delta\phi_2)^{1/2}\,{\rm d}x$, rather than the coordinate $x$, all coefficients contribute to the potential in higher-order corrections from
\begin{equation}
    \frac{d_2}{x^2}= \frac{d_2}{r^2} \left(x^{-1}\int (\phi_2^{(0)}+\delta\phi_2)^{1/2}\,{\rm d}x\right)^2\,.
\end{equation}
The second factor depends on the $c_i$, but overall corrections that include the first factor of $d_2$ are at least of second order.

Our weak-field results therefore depend on the behavior of quantum fluctuations, indirectly to leading order in relationships between $d_2$ and the $c_i$ implied by the constraints, and directly at higher orders when the geometrical radial distance is used in the effective Newton potential. The equations we solve for these expansion coefficients are implied by covariance conditions, implemented here in a model of curved space-time. They could not have been derived from quantum mechanics alone, which would not suggest any bounds such as those found for $C$ and $\Delta p_2$, in addition to the uncertainty principle. General covariance, or the existence of a curved space-time geometry for the gravitational force of a quantum state of masses, therefore implies non-trivial conditions.

There are also characteristic effects in the line element. In particular, since $d_1=0$ while $c_1\not=0$ generically, the time and space components of the space-time metric are affected in different ways by quantum corrections. This result is in contrast to previous assumptions, for instance in \cite{QuantumSwitchSphSymm}, where Schwarzschild-like patches of space-time, with closely related time and space components, were glued to each other in order to construct a geometry suitable for superposition states. Our results imply a more precise space-time geometry that combines interrelated gravitational and quantum effects.

In this way, our solutions can be used for consistent descriptions of quantum superpositions, defined by suitable values of the moments or of the parameters $C$ and $U_0$. The setting of spherical symmetry implies that a superposition state can only be formulated for masses at the same position, defining the center of symmetry. Quantum fluctuations are therefore not given by position fluctuations, but rather indirectly by mass fluctuations: the primary quantum operator, as always in canonical gravity, is given by the spatial metric, or $\phi_2$ in the present formulation. Quantum momenta of $\phi_2$, parameterized by $C$ and $U_0$, therefore determine the fundamental fluctuations. The mass enters only indirectly via the weak-field limit of the lapse function $N$, which is coupled to $\phi_2$ and its moments by the constraints and evolution equations (or staticity conditions in the present case). These indirect implications on the mass or its superpositions are the reason why superposition geometries cannot simply be constructed from wave functions, but rather have to be derived through various consistency conditions implied by the gravitational constraints. The final result, expressed in the form of a line element, can then be analyzed by standard means, for instance by computing geodesics and proper-time intervals.

Our space-time geometry can be used not only for standard relativity analysis, such as geodesics, but also for additional quantum-information studies. For instance, there is a promising line of inquiry calculating quantum switch predictions in the context of our (and related) black hole mass superpositions. If the quantum switch experiment in \cite{QuantumSwitchSphSymm} were conducted outside one such black hole superposition, how would the uncertainty within our quantum-corrected metric components impact the quantum switch experimental output? We would go about defining the measurement events in terms of proper time, which will be subject to fluctuation effects inherited from the metric components. One implication is that measurements would have to be made far enough apart in space and time for the quantum switch order to be distinct despite the uncertainties in radial and temporal distances. Increasing these distances should improve the result of the quantum switch measurement, as long as the necessary mass entanglement is preserved, throughout the necessary time and over the necessary distances. Such a quantum switch experiment would need to be conducted repeatedly to generate sufficient statistics on the results so as to associate any variation in output with the mass (and therefore metric) uncertainty from the space-time quantum corrections.\footnote{We thank Nat\'alia M\'oller for discussions about these questions.}

\section{Conclusions}

We have derived the first fundamental description of fluctuating space-time geometries that may be interpreted as superpositions of classical black holes, or as superpositions of matter constituents in a single black hole. Our analysis here was made with the assumption of spherical symmetry. Intuitively, all superposed ingredients are therefore located at the same central point. Quantum fluctuations and superposition effects first appear in metric components of the resulting geometry. By standard relativistic analysis, such as taking the weak-field limit, they then imply indirect effects on mass superpositions in the central object or on corrections to Newton's potential.

The canonical methods used here, for both gravitational and quantum physics, made it possible to derive state properties. These include results such as new conditions on quantum fluctuations from general principles, one key example being general covariance. Further conditions follow from our assumption that our solutions are static, describing a non-rotating superposition of a black hole that has settled down after a collapse process. These conditions made it possible to derive state properties, without having to impose strong assumptions on the nature of the state, such as Gaussianity or purity. Our results, derived from leading corrections by second-order moments in the gravitational Hamiltonian, could therefore belong to either a pure or a mixed state. 

The generality of our formalism is crucial for bridging fundamental questions in quantum gravity with potentially observable implications. In particular, identifying what are to be considered fundamental gravitational degrees of freedom depends on one's approach to quantum gravity. On a basic level, wave functions or superpositions of states would be formulated for these degrees of freedom. Most of these would have to be traced out to obtain a state relevant for a given observational situation. The actual tracing process is expected to be challenging, if not impossible, to perform explicitly in any complete quantum theory of gravity that includes all possible degrees of freedom, but the outcome determines the relevant final state. We eliminate assumptions on the final state, other than that it be semiclassical in the weak-field regime but it may well be mixed. In this way, we obtained results which may be considered a universal implication of quantized space-time geometries.

The condition of general covariance not only gave us a restrictive set of equations to solve, it also made it possible to derive gravitational effects in different coordinate systems. In particular, we used isotropic coordinates in the present paper, which are relevant for the weak-field limit. Our methods were originally developed in \cite{SphSymmMoments} and evaluated in Schwarzschild-type coordinates. We have seen several results that demonstrate general agreement, in particular the behavior of quantum fluctuations as a function of the radial distance. The previous paper focused on horizon properties, while the new physical result of the present paper is an effective Newton's potential, with corrections from quantum fluctuations. We found new terms in this potential that are expected to be larger than those previously derived in perturbative quantum gravity. The general nature of our quantum states explains this difference because our effects pertain to a test mass in the curved, fluctuating space-time of a large central mass. Standard results of perturbative quantum gravity instead produce an effective potential between two test masses on a flat background.

Our conditions on state parameters imply a novel lower bound on momentum fluctuations (geometrically related to the extrinsic curvature of spacelike slices). Unlike the standard uncertainty relation of quantum mechanics, this lower bound (\ref{Ineqp}) is independent of fluctuations of the configuration variable (a metric component) conjugate to this momentum. The lower bound does, however, depend inversely on the mass of the central object. Heuristically, this bound means that momentum fluctuations cannot be arbitrarily small. A quantum black hole that is static on average, with vanishing momentum expectation values as assumed here, can therefore be viewed in a new way. We interpret the black hole as an oscillating system in a superposition of expanding and collapsing classical geometries. Implications for the stability of black holes would require an extension of our model to non-static and non-spherical geometries. In such an extension, the quantum oscillation could involve higher multipoles that could source gravitational waves.

The present results for static and spherical configurations can be used in multiple ways for further analysis, mainly as a background space-time. Several questions of current interest in relativistic quantum information theory make use of the concept of proper time, which so far has mainly been defined for a classical space-time. Our effective line elements extend this important notion to quantum backgrounds with fluctuation terms. On these backgrounds, additional systems of interest for quantum information can then be set up. One example is the analysis of quantum switch experiments described in \cite{QuantumSwitchSphSymm}. It is also possible to study the quantum effects of matter fields on our backgrounds. One could, for instance, study possible effects on their entanglement or other properties, as analyzed in \cite{VacuumEntanglement,MinkowskiSup}. Another promising avenue for exploration is to look at related effects in analog gravity \cite{AnalogSup}.

\section*{Acknowledgements}

Thank you to Nat\'alia M\'oller for discussions of connections to her quantum switch work. This work was supported in part by NSF grant PHY-2206591 and a PA NASA Space Grant.

\begin{appendix}

\section{Equations for metric corrections}
\label{app}

This appendix collects details of our solution procedure of quantum-corrected constraints.

\subsection{Equations of motion}

Calculating the equations of motion from Poisson brackets of our fields and momenta with the Hamiltonian, we obtain:
\begin{eqnarray}
    \frac{\dot{\phi}_2}{c} &=& \{\phi_2, \bar{H}[N]+H_{\phi_2}[L] \} = \frac{\delta
      \bar{H}[N]}{\delta p_2}+\frac{\delta H_{\phi_2}[L]}{\delta p_2} \\
 &=& -\frac{3 \phi _3 \left(2 p_3 \phi _2+p_2 \phi _3\right)}{\sqrt{2x \phi _2}}L-\frac{3 x \left(4 p_3 \phi _2 \phi _3+p_2 \left(8 \phi _2{}^2-\phi _3{}^2\right)\right)}{4 \sqrt{2} \left(x \phi _2\right){}^{3/2}}N
\end{eqnarray}
and
\begin{eqnarray}
    \frac{\dot{\phi}_3}{c} &=& \{\phi_3, \bar{H}[N]+H_{\phi_2}[L] \} = \frac{\delta
      \bar{H}[N]}{\delta p_3}+\frac{\delta H_{\phi_2}[L]}{\delta p_3}\\
 &=& -\frac{3 \sqrt{2}  p_2 \sqrt{x \phi _2} \phi _3}{x}L-\frac{3  \left(2 p_3 \phi _2+p_2 \phi _3\right)}{ \sqrt{2x \phi _2}}N\,.
\end{eqnarray}
These equations are identically satisfied in the static case.

In addition, we have non-trivial equations of motion
\begin{eqnarray}
 0&=&\frac{\dot{p}_2}{c}=\{p_2,\bar{H}[N]+H_{\phi_2}[L]\}= -\frac{\delta \bar{H}[N]}{\delta
   \phi_2}-\frac{\delta H_{\phi_2}[L]}{\delta \phi_2}\label{p2dotL}\\
 0&=&\frac{\dot{p}_3}{c}=\{p_3,\bar{H}[N]+H_{\phi_2}[L]\}=-\frac{\delta \bar{H}[N]}{\delta
   \phi_3}-\frac{\delta H_{\phi_2}[L]}{\delta \phi_3}
\end{eqnarray}
for static behavior. We evaluate the first equation after imposing the Hamiltonian constraint. The second equation, as seen in \cite{SphSymmMoments} can be used to derive $L$, which is not required for our purpose of finding corrections to Newton's potential. (Unlike $N$, the multiplier $L$ does not appear in metric coefficients and instead determines how fluctuations contribute to the evolution generator. This function does contribute to the equation (\ref{p2dotL}) that we will solve below, but only in terms of second or higher order in the quantum parameter $\phi_3$. As we will see, for our purposes it will be sufficient to solve (\ref{p2dot}) to first order, in which $L$ does not appear.)

\subsection{Perturbations around classical background}

Fluctuation terms in the quantum Hamiltonian constraint modify the classical solutions for $\phi_2$ and $N$. 
As a first approximation, we apply a perturbative treatment to both $\phi_2$ and $N$ around the classical solutions, $\phi_2^{(0)}$ and $N^{(0)}$:
\begin{eqnarray} \label{phi2N}
 \phi_2&=&\phi_2^{(0)}+\delta\phi_2=2x\left(1+\frac{b}{x}\right)^4+\delta\phi_2\\
 N&=&N^{(0)}+\delta N=\frac{x-b}{x+b}+\delta N\nonumber
\end{eqnarray}
As discussed earlier, we absorb the original integration constant $N_0$ in $N^{(0)}$ into the definition of the time coordinate, effectively setting $N_0=1$ here. Applying these perturbations to the Hamiltonian constraint, we can arrange the terms according to whether they contain the quantum field $\phi_3$ and its momentum $p_3$ or only the classical fields, taking into account  the perturbed portions for $\phi_2$ in the latter case. We write
\begin{eqnarray} \label{Hbarx}
  \bar{H}[N]&=& H[N]+H_2[N]
\end{eqnarray}
with
\begin{eqnarray}
 H_2[N] &=& \int{\rm d}x N(x)
 \left(\frac{1}{2}\frac{\partial^2H}{\partial p_2^2}
   \left(p_3^2+\frac{U}{\phi_3^2}\right)-H_0+
   \frac{\partial^2H}{\partial\phi_2\partial p_2}\phi_3p_3\right)\nonumber\\
   &&
\left.+\frac{1}{2}\frac{\partial^2H}{\partial\phi_2^2} \phi_3^2+
   \frac{\partial^2H}{\partial\phi_2\partial\phi_2'}
   \phi_3\phi_3' +\frac{1}{2}\frac{\partial^2H}{(\partial\phi_2')^2} (\phi_3')^2+\frac{\partial^2H}{\partial\phi_2\partial\phi_2''} \phi_3\phi_3''\right)
\end{eqnarray}
and
\begin{eqnarray}
 H[N] &=& H[N]|_{\phi_2^{(0)}}
+\int{\rm d}x(N^{(0)}+\delta N)
 \left(\frac{\partial   H}{\partial\phi_2} \delta\phi_2+
\frac{\partial   H}{\partial\phi_2'} \delta\phi_2'+
\frac{\partial   H}{\partial\phi_2''} \delta\phi_2''\right)\label{Hdelta}\\
&&+ \int{\rm d}x N^{(0)} \left(
   \frac{1}{2}\frac{\partial^2H}{\partial\phi_2^2}(\delta\phi_2)^2+
   \frac{\partial^2H}{\partial\phi_2\partial\phi_2'}
   \delta\phi_2\delta\phi_2'+\frac{1}{2}
   \frac{\partial^2H}{(\partial\phi_2')^2}
   (\delta\phi_2')^2+
   \frac{\partial^2H}{\partial\phi_2\partial\phi_2''}
   \delta\phi_2\delta\phi_2''\right)\nonumber
\end{eqnarray}
where 
$H[N]|_{\phi_2^{(0)}}=0$ for classical background solutions.
We consider the quantum field $\phi_3$ and the perturbations $\delta\phi_2$ and $\delta N$ induced by it to be of the same order. Therefore, $H_2[N]$ is already of second order, such that we can simply insert the classical solutions in the coefficients defined by partial derivatives of $H$. Similarly, all non-zero correction terms in (\ref{Hdelta}) are of second order because the only linear terms (in the first line of (\ref{Hdelta}) proportional to $N^{(0)}$) are, upon integration by parts, proportional to the classical equation of motion for $p_2$, which vanishes for static solutions.

Having applied the perturbations of $N$ and $\phi_2$ to both $H[N]$ and $H_2[N]$, we now have their sum, the perturbed form of our corrected Hamiltonian $\bar{H}[N]$. We then apply the classical background solutions into the equation. At this stage, we also insert the solution for $\phi_3$, resulting in the equation 
\begin{eqnarray}
    \bar{H}[N]|_{N^{(0)},\phi_2^{(0)}} &=&-\frac{3 x^2 (b+x)^2 U}{C^2 \left(3 b^2+3 b x+x^2\right)^2}+\frac{3U}{C^2}-\frac{3 x^4 (x-b) \left(13 b^2-10 b x+x^2\right) \delta \phi _2{}^2}{16 (b+x)^9}\nonumber\\
    &&+\frac{3 C^2 \left(-24 b^6+16 b^3 (b+x)^3-(b+x)^6\right)}{16 (b+x)^8}+\frac{3 x^6 (b-x) \delta \phi _2'{}^2}{16 (b+x)^7}\nonumber\\
    &&+\frac{x^3 \delta N \delta \phi _2''}{2 (b+x)^2}+\frac{3 b x^2 \delta N \delta \phi _2'}{(b+x)^3}+\delta \phi _2 \left(\frac{x^6 (b-x) \delta \phi _2''}{8 (b+x)^7}\right)\nonumber\\
    &&+\delta \phi _2 \left(\frac{3 x^5 (x-5 b) (x-b) \delta \phi _2'}{8 (b+x)^8}+\frac{3 b x (b-x) \delta N}{(b+x)^4}\right).
\end{eqnarray}
In the second term, given by the vacuum subtraction $-H_0=3U/C^2$, we have assumed that $U$ does not depend on $b$. This assumption will be sufficient for our purposes but could easily be relaxed if needed.

We need a second constraint equation, since the Hamiltonian constraint provides a single equation coupling $\delta \phi_2$ and $\delta N$. Such a condition can be obtained from the equation of motion for $p_2$ which, unlike the background equation (\ref{p2dot}), includes perturbations from $\delta\phi_2$. Since we only need one additional equation, it is sufficient to consider the equation of motion expanded to linear order in $\delta\phi_2$. Since $\phi_3$ is of the same order as $\delta\phi_2$ and $\bar{H}$ is quadratic in $\phi_3$, the relevant equation of motion is generated by the expanded $H[N]$ without the contribution from $H_2[N]$.
With 
\begin{equation}
    H_{\rm linear}=\frac{\partial   H}{\partial\phi_2} \delta\phi_2+
\frac{\partial   H}{\partial\phi_2'} \delta\phi_2'+
\frac{\partial   H}{\partial\phi_2''} \delta\phi_2''
\end{equation}
and 
\begin{equation} \label{p2dot}
 -\frac{\dot{p}_2|_{N^{(0)},\phi_2^{(0)}}}{c}= N^{(0)}\frac{\partial H}{\partial\phi_2} -
 \left(N^{(0)}\frac{\partial 
     H}{\partial\phi_2'}\right)'+\left(N^{(0)}\frac{\partial 
     H}{\partial\phi_2''}\right)''=0.
\end{equation}

we have separated the $\phi_2$, $\phi_2'$ and $\phi_2''$ derivatives that appeared in the original functional derivative by $\phi_2$, allowing us to treat them as separate variables for the purposes of these partial derivatives. We then have 
\begin{eqnarray}
&& \frac{\dot{p}_2|_{\rm linear}}{c} = -\frac{\partial H_{\rm
     linear}}{\partial\phi_2}N^{(0)}- \frac{\partial H^{(0)}}{\partial\phi_2}
 \delta N + \left(\frac{\partial H_{\rm
     linear}}{\partial\phi_2'}N^{(0)}+ \frac{\partial H^{(0)}}{\partial\phi_2'}
 \delta N\right)'\nonumber\\
&&\qquad\qquad- \left(\frac{\partial H_{\rm
     linear}}{\partial\phi_2''}N^{(0)}+ \frac{\partial H^{(0)}}{\partial\phi_2''}
 \delta N\right)''\nonumber\\
&=& -\left(\frac{\partial^2H}{\partial\phi_2^2} 
   \delta\phi_2 + \frac{\partial^2H}{\partial\phi_2'\partial\phi_2} 
   \delta\phi_2'+ \frac{\partial^2H}{\partial\phi_2''\partial\phi_2} 
   \delta\phi_2''\right.\nonumber\\
   &&\left.- \left(\frac{\partial^2H}{\partial\phi_2'\partial\phi_2} 
   \delta\phi_2+\frac{\partial^2H}{(\partial\phi_2')^2} 
   \delta\phi_2'\right)'-\left(\frac{\partial^2H}{\partial\phi_2''\partial\phi_2} \delta\phi_2)\right)''\right)N^{(0)}\nonumber\\
&&+ \left(\frac{\partial^2H}{\partial\phi_2'\partial\phi_2} 
   \delta\phi_2+\frac{\partial^2H}{(\partial\phi_2')^2} 
   \delta\phi_2'-2\left(\frac{\partial^2H}{\partial\phi_2''\partial\phi_2}\delta\phi_2\right)'\right) N^{(0)}{}' -\frac{\partial^2H}{\partial\phi_2''\partial\phi_2}\delta\phi_2 N^{(0)}{}''\nonumber\\
   &&-\left(\frac{\partial H}{\partial\phi_2}-
     \left(\frac{\partial H}{\partial \phi_2'}\right)'+\left(\frac{\partial H}{\partial\phi_2''}\right)''\right) \delta N+
   \left(\frac{\partial H}{\partial\phi_2'}- 2\left(\frac{\partial H}{\partial\phi_2''}\right)''\right) \delta N'- \frac{\partial H}{\partial\phi_2''} \delta N''\nonumber\\
&=&-\left(\left(\frac{\partial^2H}{\partial\phi_2^2} 
   -\left(\frac{\partial^2H}{\partial\phi_2'\partial\phi_2}\right)'-\left(\frac{\partial^2H}{\partial\phi_2''\partial\phi_2}\right)''\right)
   \delta\phi_2-\left(\left(\frac{\partial^2H}{(\partial\phi_2')^2}\right)'+2\left(\frac{\partial^2H}{\partial\phi_2''\partial\phi_2}\right)'\right)\delta\phi_2'\right.\nonumber\\
   &&\left.- \frac{\partial^2H}{(\partial\phi_2')^2} 
   \delta\phi_2''\right)N^{(0)}\nonumber\\
&&+ \left(\left(\frac{\partial^2H}{\partial\phi_2'\partial\phi_2}-2\left(\frac{\partial^2H}{\partial\phi_2''\partial\phi_2}\right)'\right) 
   \delta\phi_2+\left(\frac{\partial^2H}{(\partial\phi_2')^2}-2\frac{\partial^2H}{\partial\phi_2''\partial\phi_2}\right) 
   \delta\phi_2'\right) N^{(0)}{}'\nonumber\\
   &&-\frac{\partial^2H}{\partial\phi_2''\partial\phi_2}\delta\phi_2 N^{(0)}{}''\nonumber\\
   &&-\left(\frac{\partial H}{\partial\phi_2}-
     \left(\frac{\partial H}{\partial \phi_2'}\right)'+\left(\frac{\partial H}{\partial\phi_2''}\right)''\right) \delta N+
   \left(\frac{\partial H}{\partial\phi_2'}- 2\left(\frac{\partial H}{\partial\phi_2''}\right)''\right) \delta N'- \frac{\partial H}{\partial\phi_2''} \delta N''
\end{eqnarray}
Applying the classical background solutions and the solution for $\phi_3$, we obtain:
\begin{eqnarray}
    \frac{\dot{p}_2|_{N^{(0)},\phi_2^{(0)}}}{c} &=& \frac{b x^4 \left(33 b^2-51 b x+16 x^2\right) \delta \phi _2}{4 (b+x)^9}+\frac{3 x^6 (b-x) \delta \phi _2''}{8 (b+x)^7}\\
   &&+\frac{b x^5 (15 b-16 x) \delta \phi _2'}{4 (b+x)^8}-\frac{x^3 \delta N''}{2 (b+x)^2}+\frac{3 b x \left(b (x-2)+x^2\right) \delta N'}{(b+x)^4}.\nonumber
\end{eqnarray}

\subsection{Asymptotic expansion}
\label{s:Asymp}

In order to solve this system, we first expand $\delta N$ and $\delta\phi_2$ as power series, similar to our earlier handling of $\phi_3$:  $\delta N= d_1/x+d_2/x^2+d_3/x^3$ and $\delta\phi_2=c_1+c_2/x+c_3/x^2$. The classical solution for $N$ is asymptotically constant, while the classical $\phi_2$ grows like $x$. Our perturbations include only terms higher than the classical order in $1/x$ in order to preserve the Minkowski limit.

The remaining undetermined function is $U(x)$. It is not subject to a constraint or equation of motion, but its dependence on $x$ turns out to be restricted by solvability conditions on the expansion coefficients $c_i$ and $d_i$. For $U(x)$ to be compatible with the weak-field limit, it must have an expansion in $1/x$. For simplicity, we only consider power-law forms $U(x)=U_0x^{-\kappa}$, in which the exponent is an integer $\kappa\geq0$. We will first provide detailed expressions of the constraint and equation of motion, both expanded in $y=b/x$ around zero, for the case of $\kappa=1$ and then show why the remaining power laws do not result in non-trivial real solutions.

The expanded Hamiltonian constraint is given by
\begin{eqnarray}
    \bar{H}[N] &=&y^2 \left(-\frac{3 c_1^2}{16 b^2}-\frac{3 C^2}{16 b^2}+\frac{12 U_0}{b C^2}\right)\nonumber\\
   &&+y^3 \left(\frac{c_2 d_1}{b^3}-\frac{c_1 c_2}{b^3}-\frac{3 c_1 d_1}{b^2}+\frac{15 c_1^2}{4 b^2}+\frac{3 C^2}{8 b^2}-\frac{30 U_0}{b C^2}\right)\nonumber\\
   &&+y^4 \left(\frac{3 c_3 d_1}{b^4}+\frac{c_2 d_2}{b^4}-\frac{c_2^2}{b^4}-\frac{15 c_1 c_3}{8 b^4}-\frac{8 c_2 d_1}{b^3}-\frac{3 c_1 d_2}{b^3}+\frac{59 c_1 c_2}{4 b^3}+\frac{15 c_1 d_1}{b^2}\right)\nonumber\\
   &&   +y^4 \left(-\frac{501 c_1^2}{16 b^2}-\frac{9 C^2}{16 b^2}+\frac{54 U_0}{b C^2}\right) +O(y^{5}) \label{HExpand}
\end{eqnarray}
This equation can be simplified because the constraint has to vanish for any lapse function. Instead of using the expanded lapse with coefficients $d_i$, a simpler expression is obtained for $N=N^{(0)}$. The resulting constraint can simply be obtained by setting $d_1=0$ and $d_2=0$ in the preceding expression:

\begin{eqnarray}
    \bar{H}[N^{(0)}] &=&y^2 \left(-\frac{3 c_1^2}{16 b^2}-\frac{3 C^2}{16 b^2}+\frac{12 U_0}{b C^2}\right)+y^3 \left(-\frac{c_1 c_2}{b^3}+\frac{15 c_1^2}{4 b^2}+\frac{3 C^2}{8 b^2}-\frac{30 U_0}{b C^2}\right)\nonumber\\
   &&+y^4 \left(-\frac{c_2^2}{b^4}-\frac{15 c_1 c_3}{8 b^4}+\frac{59 c_1 c_2}{4 b^3}-\frac{501 c_1^2}{16 b^2}-\frac{9 C^2}{16 b^2}+\frac{54 U_0}{b C^2}\right) +O(y^{5})
\end{eqnarray}
and we have the equation of motion
\begin{eqnarray}
    \frac{\dot{p}_2}{c} &=& -\frac{d_1 y^2}{b^2}+y^3 \left(-\frac{3 c_2}{4 b^3}-\frac{3 d_2}{b^3}+\frac{4 c_1}{b^2}-\frac{d_1}{b^2}\right)+y^4 \left(-\frac{9 c_3}{4 b^4}-\frac{6 d_3}{b^4}+\frac{14 c_2}{b^3}-\frac{195 c_1}{4 b^2}+\frac{6 d_1}{b^2}\right)\nonumber\\
   && +O(y^{5}), \label{p2dotExpand}
\end{eqnarray}
after gathering terms by powers of $y=b/x$. We then equate the second, third, and fourth order coefficients to zero for both $\bar{H}[N]=0$ and $\dot{p}_2=0$. This produces a system of six equations, which we can solve for the coefficients in each power series. For instance, the quadratic term in $y$ of the Hamiltonian constraint has real solutions from a quadratic equation for $c_1$ since $U_0>0$ by our positivity condition. These solutions then successively determine $c_2$, $d_2$, $c_3$ and $d_3$ from the remaining terms, while $d_1=0$ from the quadratic term in $\dot{p}_2/c=0$. These results are used and discussed in the main part of our paper.

Direct evaluation of the resulting Hamiltonian constraint for constant $U$ or $U\propto 1/x^2$ show that non-trivial real solutions cannot be obtained in these cases. For constant $U$, we have
\begin{eqnarray}
    \bar{H}[N] &=&\frac{12 U y}{C^2} + y^2 \left(-\frac{3 c_1^2}{16 b^2}-\frac{3 C^2}{16 b^2}-\frac{30 U}{C^2}\right)\nonumber\\
   &&+y^3 \left(\frac{c_2 d_1}{b^3}-\frac{c_1 c_2}{b^3}-\frac{3 c_1 d_1}{b^2}+\frac{15 c_1^2}{4 b^2}+\frac{3 C^2}{8 b^2}+\frac{54 U}{C^2}\right)\nonumber\\
   &&+y^4 \left(\frac{3 c_3 d_1}{b^4}+\frac{c_2 d_2}{b^4}-\frac{c_2^2}{b^4}-\frac{15 c_1 c_3}{8 b^4}-\frac{8 c_2 d_1}{b^3}-\frac{3 c_1 d_2}{b^3}+\frac{59 c_1 c_2}{4 b^3}+\frac{15 c_1 d_1}{b^2}-\frac{501 c_1^2}{16 b^2}\right)\nonumber\\
   && +y^4 \left(-\frac{9 C^2}{16 b^2}-\frac{63 U}{C^2}\right)+O(y^{5}) 
\end{eqnarray}
where a constant contribution $-3U/C^2$  has been removed by our general vacuum subtraction. However, a linear term in $y$ now remains, which as a part of the Hamiltonian constraint requires $U=0$. The second-order term in $y$ then implies imaginary solutions for $c_1$ if $C\not=0$, making the remaining coefficients complex as well. (If we choose $C=0$, we have a real $c_1=0$ but also remove all quantum corrections.)

For $U\propto 1/x^2$, we obtain
\begin{eqnarray}
    \bar{H}[N] &=&y^2 \left(-\frac{3 c_1^2}{16 b^2}-\frac{3 C^2}{16 b^2}\right)+y^3 \left(\frac{c_2 d_1}{b^3}-\frac{c_1 c_2}{b^3}-\frac{3 c_1 d_1}{b^2}+\frac{15 c_1^2}{4 b^2}+\frac{12 U_0}{b^2 C^2}+\frac{3 C^2}{8 b^2}\right)\nonumber\\
   &&+y^4 \left(\frac{3 c_3 d_1}{b^4}+\frac{c_2 d_2}{b^4}-\frac{c_2^2}{b^4}-\frac{15 c_1 c_3}{8 b^4}-\frac{8 c_2 d_1}{b^3}-\frac{3 c_1 d_2}{b^3}+\frac{59 c_1 c_2}{4 b^3}+\frac{15 c_1 d_1}{b^2}-\frac{501 c_1^2}{16 b^2}\right)\nonumber\\
   &&+y^4 \left(-\frac{30 U_0}{b^2 C^2}-\frac{9 C^2}{16 b^2}\right)+O(y^{5})
\end{eqnarray}
after removing a second-order term in $y$ as per our vacuum subtraction. The remaining quadratic term implies imaginary $c_1$, and complex values for the remaining coefficients. This behavior remains the same if we use larger exponents $\kappa>2$ because the first non-zero $U$-term then appears in higher-order contributions in $y$ and do not change the imaginary nature of the resulting $c_1$. As a power law, the form $U(x)=U_0/x$ is therefore uniquely determined by solvability conditions.

\end{appendix}

%\bibliographystyle{../preprint}
%\bibliography{../Bib/QuantGra,../Bib/Tunneling}

\begin{thebibliography}{10}

\bibitem{ProperTimeInterfer}
M.\ Zych, F.\ Costa, I.\ Pikovski, and C.\ Brukner,
\newblock Quantum interferometric visibility as a witness of general
  relativistic proper time,
\newblock {\em Nat.\ Commun.} 2 (2011) 505, [arXiv:1105.4531]

\bibitem{ProperTimeInterfer2}
M.\ Zych, F.\ Costa, I.\ Pikovski, T.~C.\ Ralph, and C.\ Brukner,
\newblock General relativistic effects in quantum interference of photons,
\newblock {\em Class.\ Quantum Grav.} 29 (2012) 224010, [arXiv:1206.0965]

\bibitem{ClassQuantProperTime}
A.~R.~H.\ Smith and M.\ Ahmadi,
\newblock Quantum clocks observe classical and quantum time dilation,
\newblock {\em Nat.\ Commun.} 11 (2020) 5360, [arXiv:1904.12390]

\bibitem{QuantumSwitchEarth}
N.~S.\ M\'oller, B.\ Sahdo, and N.\ Yokomizo,
\newblock Quantum switch in the gravity of Earth,
\newblock {\em Phys.\ Rev.\ A} 104 (2021) 042414, [arXiv:2012.03989]

\bibitem{BHMassSup}
J.\ Foo, C.~S.\ Arabaci, M.\ Zych, and R.~B.\ Mann,
\newblock Quantum signatures of black hole mass superpositions,
\newblock {\em Phys.\ Rev.\ Lett.} 129 (2022) 181301, [arXiv:2111.13315]

\bibitem{QuantumSwitchSphSymm}
N.~S.\ M\'oller, B.\ Sahdo, and N.\ Yokomizo,
\newblock Gravitational quantum switch on a superposition of spherical shells,
\newblock {\em Quantum} 8 (2024) 1248, [arXiv:2306.10984]

\bibitem{QuantumSupSpaceTime}
F.\ Giacomini and C.\ Brukner,
\newblock Quantum superposition of spacetimes obeys Einstein’s equivalence
  principle,
\newblock {\em AVS Quantum Sci.} 4 (2022) 015601, [arXiv:2109.01405]

\bibitem{QuantumSupFalling}
A.-C.\ de~la Hamette, V.\ Kabel, E.\ Castro-Ruiz, and C.\ Brukner,
\newblock Falling through masses in superposition: quantum reference frames for
  indefinite metrics,
\newblock {\em Commun.\ Phys.} 6 (2023) 231, [arXiv:2112.11473]

\bibitem{QuantumSupConformal}
V.\ Kabel, A.-C.\ de~la Hamette, E.\ Castro-Ruiz, and C.\ Brukner,
\newblock Quantum conformal symmetries for spacetimes in superposition,
  [arXiv:2207.00021]

\bibitem{ScalarFieldCurv}
M.\ Saravini, S.\ Aslanbeigi, and A.\ Kempf,
\newblock Spacetime curvature in terms of scalar field propagators,
\newblock {\em Phys.\ Rev.\ D} 93 (2016) 045026, [arXiv:1510.02725]

\bibitem{DistCorr}
A.\ Kempf,
\newblock Replacing the notion of spacetime distance by the notion of
  correlation,
\newblock {\em Front.\ Phys.} 9 (2021) 655857, [arXiv:2110.08278]

\bibitem{EffectiveNewton}
J.~F.\ Donoghue,
\newblock Leading Quantum Correction to the Newtonian Potential,
\newblock {\em Phys.\ Rev.\ Lett.} 72 (1994) 2996--2999, [gr-qc/9310024]

\bibitem{EffectiveGR}
J.~F.\ Donoghue,
\newblock General relativity as an effective field theory: The leading quantum
  corrections,
\newblock {\em Phys.\ Rev.\ D} 50 (1994) 3874--3888, [gr-qc/9405057]

\bibitem{BurgessLivRev}
C.~P.\ Burgess,
\newblock Quantum Gravity in Everyday Life: General Relativity as an Effective
  Field Theory,
\newblock {\em Living Rev.\ Relativity} 7 (2004) 5, [gr-qc/0311082],
\newblock http://www.livingreviews.org/lrr-2004-5

\bibitem{EffAcQM}
F.\ Cametti, G.\ Jona-Lasinio, C.\ Presilla, and F.\ Toninelli,
\newblock Comparison between quantum and classical dynamics in the effective
  action formalism,
\newblock In {\em Proceedings of the International School of Physics ``Enrico
  Fermi'', Course CXLIII}, pages 431--448, Amsterdam, 2000. IOS Press,
  [quant-ph/9910065]

\bibitem{EffAc}
M.\ Bojowald and A.\ Skirzewski,
\newblock Effective Equations of Motion for Quantum Systems,
\newblock {\em Rev.\ Math.\ Phys.} 18 (2006) 713--745, [math-ph/0511043]

\bibitem{Karpacz}
M.\ Bojowald and A.\ Skirzewski,
\newblock Quantum Gravity and Higher Curvature Actions,
\newblock {\em Int.\ J.\ Geom.\ Meth.\ Mod.\ Phys.} 4 (2007) 25--52,
  [hep-th/0606232],
\newblock Proceedings of ``Current Mathematical Topics in Gravitation and
  Cosmology'' (42nd Karpacz Winter School of Theoretical Physics), Ed.\
  Borowiec, A.\ and Francaviglia, M.

\bibitem{SphSymmMoments}
K.\ Berglund, M.\ Bojowald, M.\ D\'{\i}az, and G.\ Sims,
\newblock Quasiclassical solutions for static quantum black holes,
\newblock {\em Phys.\ Rev.\ D} 109 (2024) 024006, [arXiv:2012.07649]

\bibitem{DiracHamGR}
P.~A.~M.\ Dirac,
\newblock The theory of gravitation in Hamiltonian form,
\newblock {\em Proc.\ Roy.\ Soc.\ A} 246 (1958) 333--343

\bibitem{Katz}
J.\ Katz,
\newblock Les crochets de Poisson des contraintes du champ gravitationne,
\newblock {\em Comptes Rendus Acad.\ Sci.\ Paris} 254 (1962) 1386--1387

\bibitem{ADM}
R.\ Arnowitt, S.\ Deser, and C.~W.\ Misner,
\newblock The Dynamics of General Relativity, In L.\ Witten, editor, {\em
  Gravitation: An Introduction to Current Research},
\newblock Wiley, New York, 1962,
\newblock Reprinted in \cite{ADMRe}

\bibitem{Regained}
S.~A.\ Hojman, K.\ Kucha\v{r}, and C.\ Teitelboim,
\newblock Geometrodynamics Regained,
\newblock {\em Ann.\ Phys.\ (New York)} 96 (1976) 88--135

\bibitem{CUP}
M.\ Bojowald,
\newblock {\em Canonical Gravity and Applications: Cosmology, Black Holes, and
  Quantum Gravity},
\newblock Cambridge University Press, Cambridge, 2010

\bibitem{Foundations}
M.\ Bojowald,
\newblock {\em Foundations of Quantum Cosmology},
\newblock IOP Publishing, London, UK, 2020

\bibitem{HigherMoments}
M.\ Bojowald, D.\ Brizuela, H.~H.\ Hernandez, M.~J.\ Koop, and H.~A.\
  Morales-T\'ecotl,
\newblock High-order quantum back-reaction and quantum cosmology with a
  positive cosmological constant,
\newblock {\em Phys.\ Rev.\ D} 84 (2011) 043514, [arXiv:1011.3022]

\bibitem{Arnold}
V.~I.\ Arnold,
\newblock {\em Mathematical Methods of Classical Mechanics},
\newblock Springer, 1997

\bibitem{Weinstein}
A.\ Cannas~da Silva and A.\ Weinstein,
\newblock {\em Geometric models for noncommutative algebras}, volume~10 of {\em
  Berkeley Mathematics Lectures},
\newblock Am.\ Math.\ Soc., Providence, 1999

\bibitem{VariationalEffAc}
R.\ Jackiw and A.\ Kerman,
\newblock Time Dependent Variational Principle And The Effective Action,
\newblock {\em Phys.\ Lett.\ A} 71 (1979) 158--162

\bibitem{EnvQuantumChaos}
R.~A.\ Jalabert and H.~M.\ Pastawski,
\newblock Environment-independent decoherence rate in classically chaotic
  systems,
\newblock {\em Phys.\ Rev.\ Lett.} 86 (2001) 2490--2493

\bibitem{QHDTunneling}
O.\ Prezhdo,
\newblock Quantized Hamiltonian Dynamics,
\newblock {\em Theor.\ Chem.\ Acc.} 116 (2006) 206

\bibitem{CQC}
T.\ Vachaspati and G.\ Zahariade,
\newblock A Classical-Quantum Correspondence and Backreaction,
\newblock {\em Phys.\ Rev.\ D} 98 (2018) 065002, [arXiv:1806.05196]

\bibitem{Bosonize}
B.\ Bayta\c{s}, M.\ Bojowald, and S.\ Crowe,
\newblock Faithful realizations of semiclassical truncations,
\newblock {\em Ann.\ Phys.} 420 (2020) 168247, [arXiv:1810.12127]

\bibitem{EffPotRealize}
B.\ Bayta\c{s}, M.\ Bojowald, and S.\ Crowe,
\newblock Effective potentials from canonical realizations of semiclassical
  truncations,
\newblock {\em Phys.\ Rev.\ A} 99 (2019) 042114, [arXiv:1811.00505]

\bibitem{EffCons}
M.\ Bojowald, B.\ Sandh\"ofer, A.\ Skirzewski, and A.\ Tsobanjan,
\newblock Effective constraints for quantum systems,
\newblock {\em Rev.\ Math.\ Phys.} 21 (2009) 111--154, [arXiv:0804.3365]

\bibitem{EffConsRel}
M.\ Bojowald and A.\ Tsobanjan,
\newblock Effective constraints for relativistic quantum systems,
\newblock {\em Phys.\ Rev.\ D} 80 (2009) 125008, [arXiv:0906.1772]

\bibitem{EffConsQBR}
M.\ Bojowald and S.\ Brahma,
\newblock Effective constraint algebras with structure functions,
\newblock {\em J.\ Phys.\ A: Math.\ Theor.} 49 (2016) 125301, [arXiv:1407.4444]

\bibitem{DirQuant}
P.~A.~M.\ Dirac,
\newblock {\em Lectures on Quantum Mechanics},
\newblock Yeshiva Press, 1969

\bibitem{LongRangeQG}
I.~J.\ Muzinich and S.\ Vokos,
\newblock Long Range Forces in Quantum Gravity,
\newblock {\em Phys.\ Rev.\ D} 52 (1995) 3472--3483, [hep-th/9501083]

\bibitem{QGNewton}
H.~W.\ Hamber and S.\ Liu,
\newblock On the Quantum Corrections to the Newtonian Potential,
\newblock {\em Phys.\ Lett.\ B} 357 (1995) 51--56, [hep-th/9505182]

\bibitem{NewtonOneLoop}
A.\ Akhundov, S.\ Bellucci, and A.\ Shiekh,
\newblock Gravitational interaction to one loop in effective quantum gravity,
\newblock {\em Phys.\ Lett.\ B} 395 (1997) 16--23, [gr-qc/9611018]

\bibitem{TwoMasses}
N.~E.~J.\ Bjerrum-Bohr, J.~F.\ Donoghue, and B.~R.\ Holstein,
\newblock Quantum Gravitational Corrections to the Nonrelativistic Scattering
  Potential of Two Masses,
\newblock {\em Phys.\ Rev.\ D} 67 (2003) 084033, [hep-th/0211072]

\bibitem{NewtonKerr}
N.~E.~J.\ Bjerrum-Bohr, J.~F.\ Donoghue, and B.~R.\ Holstein,
\newblock Quantum Corrections to the Schwarzschild and Kerr Metrics,
\newblock {\em Phys.\ Rev.\ D} 68 (2003) 084005, [hep-th/0211071]

\bibitem{EffNewtonCoulomb}
S.\ Faller,
\newblock Effective Field Theory of Gravity: Leading Quantum Gravitational
  Corrections to Newtons and Coulombs Law,
\newblock {\em Phys.\ Rev.\ D} 77 (2008) 124039, [arXiv:0708.1701]

\bibitem{VacuumEntanglement}
E.\ Martin-Martinez, A.~R.~H.\ Smith, and D.~R.\ Terno,
\newblock Spacetime structure and vacuum entanglement,
\newblock {\em Phys.\ Rev.\ D} 93 (2016) 044001, [arXiv:1507.02688]

\bibitem{MinkowskiSup}
J.\ Foo, C.~S.\ Arabaci, M.\ Zych, and R.~B.\ Mann,
\newblock Quantum superpositions of Minkowski spacetime,
\newblock {\em Phys.\ Rev.\ D} 107 (2023) 045014, [arXiv:2208.12083]

\bibitem{AnalogSup}
C.\ Barcel\'o, L.~J.\ Garay, and G.\ Garc\'{\i}a-Moreno,
\newblock Analogue gravity simulation of superpositions of spacetimes,
\newblock {\em Eur.\ Phys.\ J.\ C} 82 (2022) 727, [arXiv:2104.15078]

\bibitem{ADMRe}
R.\ Arnowitt, S.\ Deser, and C.~W.\ Misner,
\newblock The Dynamics of General Relativity,
\newblock {\em Gen.\ Rel.\ Grav.} 40 (2008) 1997--2027

\end{thebibliography}

\end{document}